\documentclass[journal]{IEEEtran}
\usepackage{amsbsy}
\usepackage{graphics}
\usepackage{graphicx}
\usepackage{amsfonts}
\usepackage{amsmath}
\usepackage{amssymb}
\usepackage{fancyhdr}
\usepackage{color}

\DeclareMathOperator{\Var}{Var} \DeclareMathOperator{\Ave}{Ave}

\newtheorem{lemma}{Lemma}
\newtheorem{Theorem}{Theorem}

\newcommand{\eqd}{\,{\buildrel d \over =}\,}

\graphicspath{{grayfigs/}}

\renewcommand{\baselinestretch}{2}
\onecolumn

\begin{document}
\large
\title{Target Detection via Network Filtering}
\author{Shu Yang and Eric D. Kolaczyk\thanks{Shu Yang and Eric D. Kolaczyk
are with the Department of Mathematics and Statistics, Boston
University, Boston, MA 02215 USA (email: shuyang@math.bu.edu;
kolaczyk@math.bu.edu).  This work was supported in part by NIH award
GM078987.}, {\it Senior Member, IEEE}}

\maketitle

\begin{abstract}
A method of `network filtering' has been proposed recently to detect
the effects of certain external perturbations on the interacting
members in a network. However, with large networks, the goal of
detection seems a priori difficult to achieve, especially since the
number of observations available often is much smaller than the
number of variables describing the effects of the underlying
network. Under the assumption that the network possesses a certain
sparsity property, we provide a formal characterization of the
accuracy with which the external effects can be detected, using a
network filtering system that combines Lasso regression in a sparse
simultaneous equation model with simple residual analysis. We
explore the implications of the technical conditions underlying our
characterization, in the context of various network topologies, and
we illustrate our method using simulated data.
\end{abstract}

\section{Introduction}
%
%
%
%
\label{sec:intro}

\begin{figure}
\begin{center}
\includegraphics[scale = 0.5]{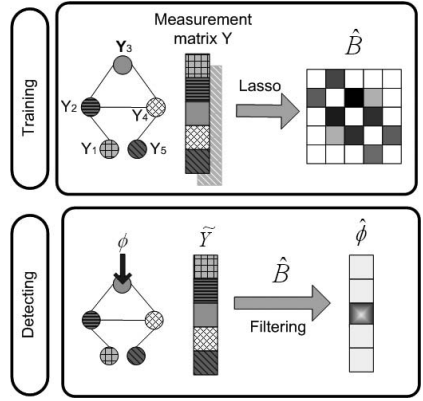}
\end{center}
\caption{Schematic illustration of the network filtering process
    proposed in this paper, shown in two stages.  In the first stage,
    the aim is to recover information on the correlation (i.e., $\hat B$)
    among the five network units, given training data $Y$.
    In the second stage, that information is used to filter new data
    $\tilde Y$, produced in the presence of an effect external to the
    system (i.e., $\phi$), so as to detect the target of that
    effect.
\label{fig:illus}}
\end{figure}

A canonical problem in statistical signal and image processing is
the detection of localized targets against complex backgrounds,
which often is likened to the proverbial task of `finding a needle
in a haystack'. In this paper, we consider the task of detecting
such targets when the `background' is neither a one-dimensional
signal nor a two-dimensional image, but rather consists of the
`typical' behavior of interacting units in a network system.  More
specifically, we assume network-indexed data, where measurements are
made on each of the units in the system and the interaction among
these units manifests itself through the correlations among these
measurements. Then, given the possible presence of an external
effect applied to a unit(s) of this system, we take as our goal the
task of identifying the location and magnitude of this effect. It is
expected that evidence of this effect be diffused throughout the
system, to an extent determined by the underlying network of
interactions among system units, like the blurring of a point source
in an image.  As a result, an appropriate filtering of the observed
measurements is necessary.  These ideas are illustrated
schematically in Figure~\ref{fig:illus}.

While networks have been an important topic of study for some time,
in recent years there has been an enormous surge in interest in the
topic, across various diverse areas of science. Examples include
computer traffic networks (e.g., \cite{crov.bala}), biological
networks (e.g., \cite{alon.sys.bio}), social networks (e.g.,
\cite{wasserman.faust}), and sensor networks (e.g.,
\cite{sensor.nets}).  Our network filtering problem was formulated
by, and is largely motivated by the work of, Cosgrove {\it et
al.}\cite{Elissa}, who used it to tackle the problem of predicting
genetic targets of biochemical compounds proposed as candidates for
drug development.  However, the problem is clearly general and it is
easy to conceive of applications in other domains.

The authors in~\cite{Elissa} model the acquisition of network data,
including the potential presence of targets, using a system of
sparse simultaneous equation models (SSEMs), and propose to search
for targets using a simple two-step procedure.  In the first step,
sparse statistical inference techniques are used to remove the
`background' of network effects, while in the second step, outlier
detection methods are applied to the resulting network-filtered
data. Empirical work presented in~\cite{Elissa}, using both
simulated data and real data from micro-array studies, demonstrates
that such network filtering can be highly effective.  However, there
is no accompanying theoretical work in~\cite{Elissa}.

In this paper, we present a formal characterization of the
performance of network filtering, exploring under what conditions
the methodology can be expected to work well.  A collection of
theoretical results are provided, which in turn are supported by an
extensive numerical study.  Particular attention is paid to the
question of how network structure influences our ability to detect
external effects.  The technical aspects of our work draw on a
combination of tools from the literatures on sparse regression and
compressed sensing, complex networks, and spectral graph theory.

The remainder of the paper is organized as follows. The basic SSEM
model and two-step network filtering methodology are presented
formally in Section~\ref{sec:nf}.  In Section~\ref{sec:est.B} we
characterize the accuracy with which the network effects can be
learned from training data, while in Section~\ref{sec:est.phi}, we
use these results to quantify the extent to which external effects
will be evident in test data after filtering out the learned network
effects.  Numerical results, based on simulations under various
choices of network structure, are presented in
Section~\ref{sec:simul}. Finally, we conclude with some additional
discussion in Section~\ref{sec:disc}.  Proofs of all formal results
are gathered in the appendices.

\section{Network Filtering: Model and Methodology}
\label{sec:nf}

Consider a system of $p$ units (e.g., genes, people, sensors, etc.).
We will assume that we can take measurements at each unit, and that
these measurements are likely correlated with measurements at some
(small) fraction of other units, such as might occur through
`interaction' among the units.  For example, in~\cite{Elissa}, where
the units are genes, the measurements are gene expression levels
from microarray experiments.  Genes regulating other genes can be
expected to have expression profiles correlated across experiments.
Alternatively, we could envision measuring environmental variables
(e.g., temperature) at nodes in a sensor network.  Sensors located
sufficiently close to each other, with respect to the dynamics of
the environmental process of interest, can be expected to yield
correlated readings.

We will also assume that there are two possible types of
measurements: a training set, obtained under `standard' conditions,
and a test set measured under the influence of additional `external'
effects.  The training set will be used to learn patterns of
interaction among the units (i.e., our `network'), and with that
knowledge, we will seek to identify in the test data those units
targeted by the external effects.

We model these two types of measurements using systems of
simultaneous equation models (SEMs). Formally, suppose that for each
of the $p$ units, we have in the training set $n$ replicated
measurements, which are assumed to be realizations of the elements
of a random vector $Y=(Y_1,\ldots,Y_p)'$. Let $Y_i$ be the
\emph{i-th} element of $Y$, and let $Y_{[-i]}$ denote all elements
of $Y$ except $Y_i$.  We specify a conditional linear relationship
among these elements, in the form
\begin{equation}
Y_i\,\big\vert\, Y_{[-i]}=y_{[-i]} \quad
  \eqd \quad \sum_{j\neq i}^p \beta_{ij} y_j + e_i \enskip ,\label{mainReg}
\end{equation}
where $\beta_{ij}$ represents the strength of association of the
measurement for the \emph{i-th} unit with that of the \emph{j-th}
unit, and $e_i$ are error terms, assumed to be independently
distributed as $N(0, \sigma^2)$.  That is, we specify a so-called
`conditional Gaussian model' for the $Y_i$, which in turn yields a
joint distribution for $Y$ in the form
\begin{equation} {Y} \,\sim\, N\left(\,0\, ,\,
(I-B)^{-1}\sigma^2\,\right), \label{traindist}
\end{equation}
with $B$ being a matrix whose $(i,j)$ entry $B_{ij}$ is
$\beta_{ij}$, for $i\neq j$, and zero otherwise. See~\cite[Ch
6.3.2]{Cressie}.  The matrix $I-B$ is assumed to be positive
definite.  In addition, we will assume $B$ (and hence $I-B$) to be
sparse, in the sense of having a substantial proportion of its
entries equal to zero.  A more precise characterization of this
assumption is given below, in the statement of Theorem~1.

We can associate a network with this model using the framework of
graphical models (e.g., \cite{lauritzen}).  Let each unit in our
system correspond to a vertex in a vertex set $V=\{1,\ldots,p\}$,
and define an edge set $E$ such that $\{i,j\}\in E$ if and only if
$B_{ij}\ne 0$.  Then the model in (\ref{traindist}), paired with the
graph $G=(V,E)$, is a Gaussian graphical model, with concentration
(or `precision') matrix $\Omega=(I-B)\sigma^{-2}$ and concentration
graph $G$.  Since we assume $B$ to be sparse, the graph $G$ likewise
will be sparse.  Gaussian graphical models are a common choice for
modeling multivariate data with potentially complicated correlation
(and hence dependency) structure.  This structure is encoded in the
graph $G$, and questions regarding the nature of this structure
often can be rephrased in terms of questions involving the topology
of $G$.  In recent years, there has been increased interest in
modeling and inference for large, sparse Gaussian graphical models
of the type we consider here (e.g., \cite{Dobra,Meinshausen}).

For the test set, our observations are assumed to be realizations of
another random vector, say
$\tilde{Y}=(\tilde{Y}_1,\ldots,\tilde{Y}_p)'$, the elements of which
differ from those of $Y$ only through the possible presence of an
additive perturbation.  That is, we model each $\tilde Y_i$,
conditional on the others, as
\begin{equation}
\tilde{Y}_i\,\big\vert\, \tilde Y_{[-i]}=\tilde y_{[-i]}\quad \eqd
\quad \sum_{j\neq i}^p \beta_{ij} \tilde{y}_j + \phi_i + \tilde{e}_i
\enskip ,\label{test}
\end{equation}
where $\phi_i$ denotes the effect of the external perturbation for
the \emph{i-th} unit, and the error terms $\tilde{e}_i$ are again
independently distributed as $N(0, \sigma^2)$.  Similar to
(\ref{traindist}), we have in this scenario
\begin{equation}
{\tilde{Y}} \,\sim\, N\left(\,(I-B)^{-1}\phi\, ,\,
                (I-B)^{-1}\sigma^2\,\right) \enskip ,
\label{testdist}
\end{equation}
where $\phi = (\phi_1,\ldots, \phi_p)'$.

The external effects $\phi$ are assumed unknown to us but sparse.
That is, we expect only a relatively small proportion of units to be
perturbed. Our objective is to estimate the external effects $\phi$
and to detect which units $i$ were perturbed i.e., to detect those
units $i$ for which $\phi_i$ stands out from zero above noise.  But
we do not observe the external effects $\phi$ directly.  Rather,
these effects are `blurred' by the network of interactions captured
in $B$, as indicted by the expression for the mean vector in
(\ref{testdist}).  If $B$ were known, however, it would be natural
to filter the data $\tilde Y$, producing
\begin{equation}
\hat\phi^{ideal} = (I-B)\tilde Y \enskip . \label{eq:phi.hat.ideal}
\end{equation}
The random vector $\hat\phi^{ideal}$ has a multivariate Gaussian
distribution, with expectation $\phi$ and covariance
$(I-B)\sigma^2$. Hence, element-wise, each $\hat\phi^{ideal}_i$ is
distributed as $N(\phi_i,\sigma^2)$, and therefore the detection of
perturbed units $i$ reduces to detection of a sparse signal against
a uniform additive Gaussian noise, which is a well-studied problem.
Note that under this model, we expect the noise in
$\hat\phi^{ideal}$ to be correlated.  However, given the assumptions
of sparsity on $B$, these correlations will be relatively localized.

Of course, $B$ typically is not known in practice, and so
$\hat\phi^{ideal}$ in (\ref{eq:phi.hat.ideal}) is an unobtainable
ideal. Studying the same problem, Cosgrove {\it et
al.}~\cite{Elissa} proposed a two-stage procedure in which (i) $p$
simultaneous sparse regressions are performed to infer $B$,
row-by-row, yielding an estimate $\hat B$, and (ii) the ideal
residuals in (\ref{eq:phi.hat.ideal}) are predicted by the values
\begin{equation}
\hat\phi = (I-\hat B)\tilde Y \enskip , \label{eq:phi.hat}
\end{equation}
after which detection is carried out\footnote{Technically, Cosgrove
{\it et al.} work under a model that differs slightly from ours,
sharing the same conditional distributions, but arrived at through
specification of a different joint distribution.  See~\cite[Ch
6.3]{Cressie} for discussion comparing such `simultaneous Gaussian
models' with our conditional Gaussian model.}.  They dubbed this
overall process `network filtering'.  A schematic illustration of
network filtering is shown in Figure~\ref{fig:illus}.

Our central concern in this paper is with characterizing the
conditions under which network filtering can be expected to work
well.  Motivated by the original context of Cosgrove {\it et al.},
involving a network of gene interactions and measurements based on
micro-array technology, we assume here that (i) $p\gg n$, (ii) the
matrix $B$ is sparse, and (iii) the vector $\phi$ is sparse.  In
carrying out our study, we adopt a strategy for estimating $B$ based
on Lasso regression~\cite{Tib}, a now-canonical example of sparse
regression. Specifically, motivated by (\ref{mainReg}), we estimate
each row $B_{i\bullet}$ as
\begin{equation}
\hat{B}_{i\bullet} \stackrel{\triangle}{=}
\arg\min_{\beta\in\mathbb{R}^p: \beta_{ii}=0}
    \|y_{i} - \sum_{j\neq i}^p \beta_{ij} y_j\|^2_2
        + \mu\|\beta\|_1\ ,
\label{eq:lasso.est}
\end{equation}
where $\mu>0$ is a regularization parameter.  Following this
estimation stage, we carry out detection using simple rank-based
procedures.

We present our results in two stages, first describing conditions
under which $\hat B$ estimates $B$ accurately, given the system of
sparse simultaneous equation models (SSEMs) defined by
(\ref{mainReg}), and then discussing the nature of the resulting
vector $\hat\phi$.  In both stages, we explore the implications of
the topological structure of $G$ on our results.


\section{Accuracy in estimation of $B$}
\label{sec:est.B}

At first glance, accurate estimation of $B$ seems impossible, since
even if the error terms $e_i$ are small, this noise typically will
be inflated by naive inversion of our systems of equations (i.e.,
because $p\gg n$). However, recent work on analogous problems in
other models has shown that under certain conditions, and using
tools of sparse inference, it is indeed possible to obtain good
estimates.  Results of this nature have appeared under the names
`compressed sensing', `compressive sampling', and similar. See the
recent review~\cite{cs.review}, and the references therein.  The
following result is similar in spirit to these others, for the
particular sparse simultaneous equation models we study here.
\medskip

\begin{Theorem}
{\it Assume the training model defined in (\ref{mainReg}) and
(\ref{traindist}), and set $\Sigma = (I-B)^{-1}\sigma^2$.  Let $S$
be the largest number of non-zero entries in any row of $B$, and
suppose that $\Sigma, S, p,$ and $n$ satisfy the conditions
\begin{eqnarray}\frac{\lambda_{max}(\Sigma)}{\lambda_{min}(\Sigma)} &\leq&
\left(\frac{1+\sqrt{S/n}}{1-\sqrt{S/n}} \right )^2 \label{cond3} \\
\hbox{and} \hspace{1.5in} & & \nonumber \\
\rho(r) & < & 1 \enskip .
 \label{cond4}
\end{eqnarray}
Here $\lambda_{max}$ and $\lambda_{min}$ refer to maximum and
minimum eigenvalues and $\rho(r) = (1+f(4r))^2+2(1+f(5r))^2-3$,
where $r = S/(p - 1)$ and $f$ is a function to be defined
latter\footnote{Specifically, $f$ is defined in Section III,
sub-section B, immediately following equation~(\ref{eq:def.rho}).}.
Finally, assume that $\mu^2\le (C_0\sigma^2 \zeta_n^-)/S$, for a
constant $C_0=C_0(n,p,r)$ and $\zeta_n^{-} = n \left(1-4(\log_2 n /
n)^{1/2}\right)$. Let $p=n^{\nu}$, for $\nu > 1$.  Then it follows
that, with overwhelming probability, for every row $B_{i\bullet}$ of
$B$ the estimator $\hat B_{i\bullet}$ in (\ref{eq:lasso.est})
satisfies the relation
\begin{equation}
\|\hat B_{i\bullet} - B_{i\bullet}\|^2_2\leq C\sigma^2\zeta_n^{+}
\enskip , \label{eq:cs.bound}
\end{equation}
where $\zeta_n^{+} = n \left(1+4(\log_2 n / n)^{1/2}\right)$ and
$C>0$ is a constant. }
\end{Theorem}
\medskip

\noindent{\em Remark 1:} The accuracy of $\hat B_{i\bullet}$ is seen
in (\ref{eq:cs.bound}) to depend primarily on the product
$n\sigma^2$ and on $C$. The constant $C$ can be bounded by an
expression of the form $(1-\rho(r))^{-2}$ times a constant depending
only on the structure of $\Sigma$.  The magnitude of $C$ therefore
is controlled essentially by the extent to which $\rho(r)$ is less
than $1$, which in turn is a rough reflection of the sparsity of the
network. Hence, in order to have good accuracy, $\sigma^2$ must be
small compared to $n^{-1}$.  In particular, if $\sigma^2 =
O(n^{-\omega})$, for $\omega > 1$, then the error in
(\ref{eq:cs.bound}) behaves roughly like $O\left(n^{-(\omega
-1)}\right)$.
\smallskip

\noindent {\em Remark 2:} Clearly it cannot be expected that we
estimate $B$ with high accuracy in all situations.  The expressions
in (\ref{cond3}) and (\ref{cond4}) dictate sufficient conditions
under which, with overwhelming probability (meaning with probability
decaying exponentially in $p$), we can expect to do well.  Due to
the intimate connection between the covariance $\Sigma$ and the
concentration graph $G$, these conditions effectively place
restrictions on the structure of the network we seek to filter, with
(\ref{cond3}) controlling the relative magnitude of the eigenvalues
of the matrix $I-B$, and (\ref{cond4}), its sparseness. Note that
since $S$ is simply the maximum degree of $G$, condition
(\ref{cond4}) relates the maximum extent of the degree distribution
of $G$ to the sample size $n$.  We explore the nature of these
conditions in more detail immediately below.
\smallskip

\noindent {\em Remark 3.} In general, of course, choice of the Lasso
regularization parameter $\mu>0$ in (\ref{eq:lasso.est}) matters.
The statement of Theorem~1 includes constraints on the range of
acceptable values for this parameter. In particular, it suggests
that $\mu$ should vary like $\sigma^2n/S$, which for
$\sigma^2=O(n^{-\omega})$ means we want
$\mu=O(S^{-1}n^{-(\omega-1)})$. The theorem does not, however,
provide explicit guidance on how to set this parameter in practice.
For the empirical work shown later in this paper, we have used
cross-validation, which we find yields results like those predicted
by the theorem over a broad range of scenarios.
\smallskip

\noindent {\em Remark 4.} There are results in the literature that
address other problems sharing certain aspects of our network
filtering problem, but none that address all together.  For example,
the bound in (\ref{eq:cs.bound}) is like that in work by Cand\`es
and Tao and colleagues (e.g., \cite{Candes,CRT}), although for a
single regression, rather than a system of simultaneous regressions.
In addition, those authors use constrained minimization for
parameter estimation, rather than Lasso-based optimization.  As
Zhu~\cite{Zhu} has recently pointed out, there are small but
important differences in these closely related problems. Our proof
makes use of Zhu's results. Similarly, Greenshtein and
Ritov~\cite{Ritov} present results for models that -- in principle
at least -- include the individual univariate regressions in
(\ref{mainReg}), although again their results do not encompass a
system of such regressions.  Furthermore, their results are in terms
of mean-squared prediction error, rather than in terms of the
regression coefficients themselves.  Finally, Meinshausen and
B\"uhlmann~\cite{Meinshausen} have studied the use of Lasso in the
context of Gaussian graphical models, but for the purpose of
recovering the topology of $G$ i.e., for variable selection, rather
than parameter estimation. The proof of Theorem~1 may be found in
Appendix~A.
\medskip

In the remainder of this section, we examine conditions
(\ref{cond3}) and (\ref{cond4}) in greater depth. These conditions
derive from our use of certain concentration inequalities, which --
although central to the proof of our result -- can be expected to be
somewhat conservative. Our numerical results, shown later, confirm
this expectation.  Nonetheless, these conditions are useful in that
they help provide insight into the way that the network topology
structure, on the one hand, and the sample size $n$, on the other,
can be expected to interact in determining the performance of our
network filtering methodology.

\subsection{ The eigenvalue constraint}

Recall that the covariance matrix $\Sigma$ is proportional to $(I -
B)^{-1}$.  In order to better understand the condition on the
covariance matrix in (\ref{cond3}), consider the special case of
\begin{equation}
\Sigma^{-1} = (I-B) = I + qD^{-1/2}A D^{-1/2} \enskip ,
\label{eq:toy.cov}
\end{equation}
where $A$ is the adjacency matrix for a graph $G$,
$D=\hbox{diag}[(d_i)]_{i\in V}$ is a diagonal matrix, $d_i$ is the
degree (i.e., the number of neighbors) of vertex $i$, and $q>0$ is a
constant.  Here the covariance $\Sigma$ is defined entirely in terms
of the topology of the concentration graph $G$.  While later, in
Sections~3 and~4, we use simulation to explore more complicated
covariance structures, where the $B_{ij}$ are assigned randomly
according to certain distributions, the simplified form in
(\ref{eq:toy.cov}) is useful in allowing us to produce analytical
results.  In particular, conditions on $\Sigma$ reduce to conditions
on our network topology\footnote{We note that (\ref{eq:toy.cov}) can
be rewritten in the form $\Sigma^{-1} = (1+q)I - q\cdot
\mathcal{L}$, where $\mathcal{L}$ is the (normalized) Laplacian
matrix of the graph $G$.  In other words, the precision matrix
$\Omega=\Sigma^{-1}$ in this simple model is just a modified
Laplacian matrix.}. For example, the following theorem describes a
sufficient condition under which (\ref{cond3}) holds for this model.
\medskip

\begin{Theorem}
{\it Suppose that the covariance matrix $\Sigma$ from Theorem~1 is
defined through (\ref{eq:toy.cov}), with $0<q<1$.  Denote
\begin{equation}
\eta_1 = \frac{\sum_{i=1}^p \frac{\sum_{j\sim i}1/d_j}{d_i} }{p} \
\hbox{and}\ \eta_2 = \left(\frac{1+\sqrt{d_{\max}/n}}{1 -
\sqrt{d_{\max}/n}}\,\Big/\,\sqrt{2}\right)^{4/p}, \label{eq:etas}
\end{equation}
where $i\sim j$ indicates that the vertices $i$ and $j$ are
neighbors in $G$ and $d_{max}=\max_{1\le i\le p} d_i$ is the maximum
vertex degree. Then condition (\ref{cond3}) on the eigenvalues of
$\Sigma$ is satisfied if
\begin{equation}
\frac{1}{(1+q)^2} + (\frac{q}{1+q})^2\, \eta_1  \leq \eta_2\enskip .
\label{eq:eta.cond}
\end{equation}
}
\end{Theorem}
\medskip

Proof of this result may be found in Appendix~B.  The restriction on
$q$ ensures that the matrix $\Sigma^{-1}$ is diagonally dominant,
which is needed for our proof, although it likely could be weakened.
Note that the condition in (\ref{eq:eta.cond}) involves the graph
$G$ only through the degree sequence $\{d_1,\ldots,d_p\}$.  More
precisely, this condition relates the average harmonic mean of
neighbor degrees (i.e., $\eta_1$) and the maximum degree to the
sample size $n$ and the constant $q$.  Accordingly, given a network,
it is straightforward to explore the implications of this condition
numerically. For example, we can explore the range of values $q$ for
which the condition holds, given $n$.

 \begin{figure}
     \begin{center}
     \includegraphics[scale = 0.2,clip = true]{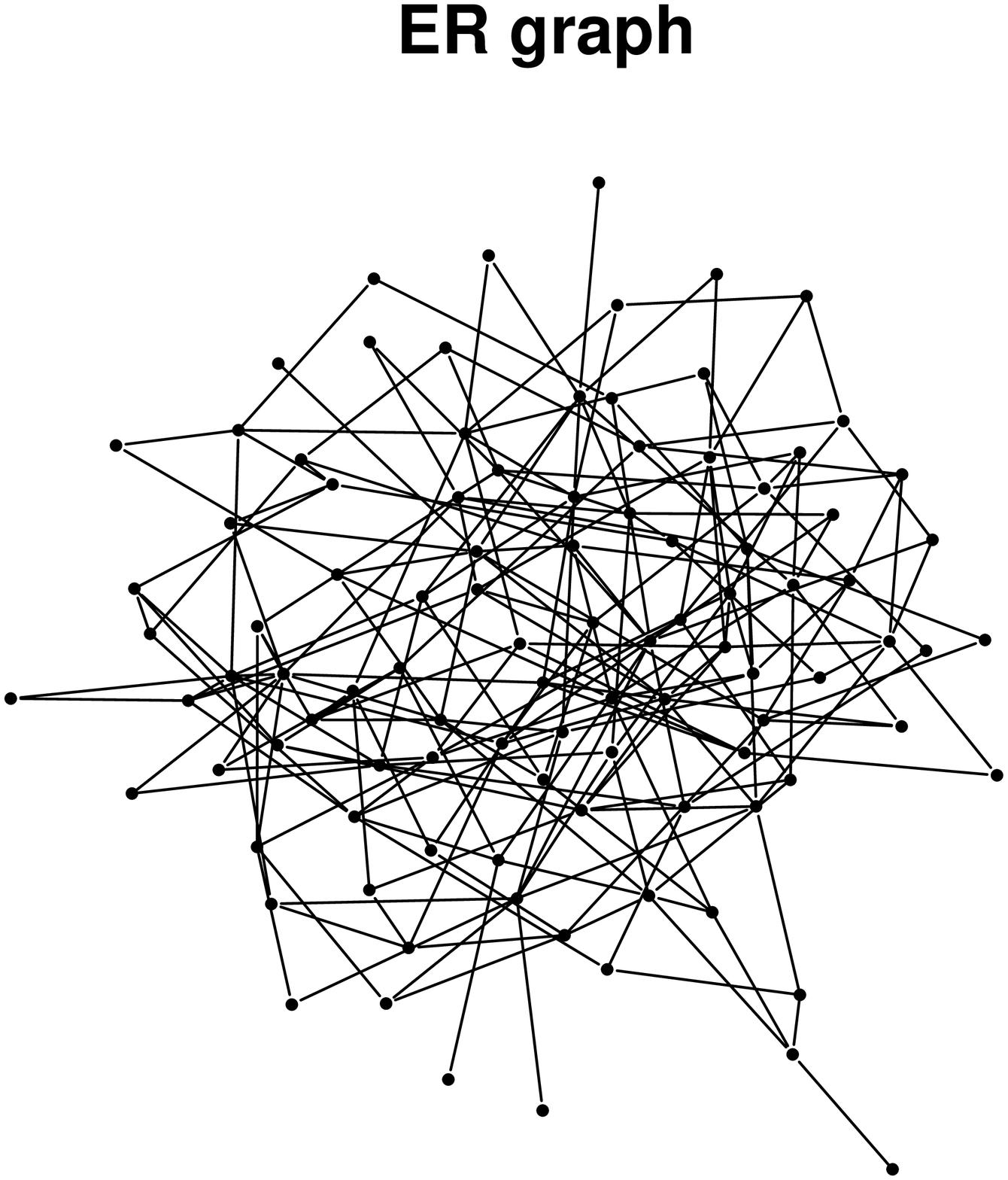}
     \includegraphics[scale = 0.2,clip = true]{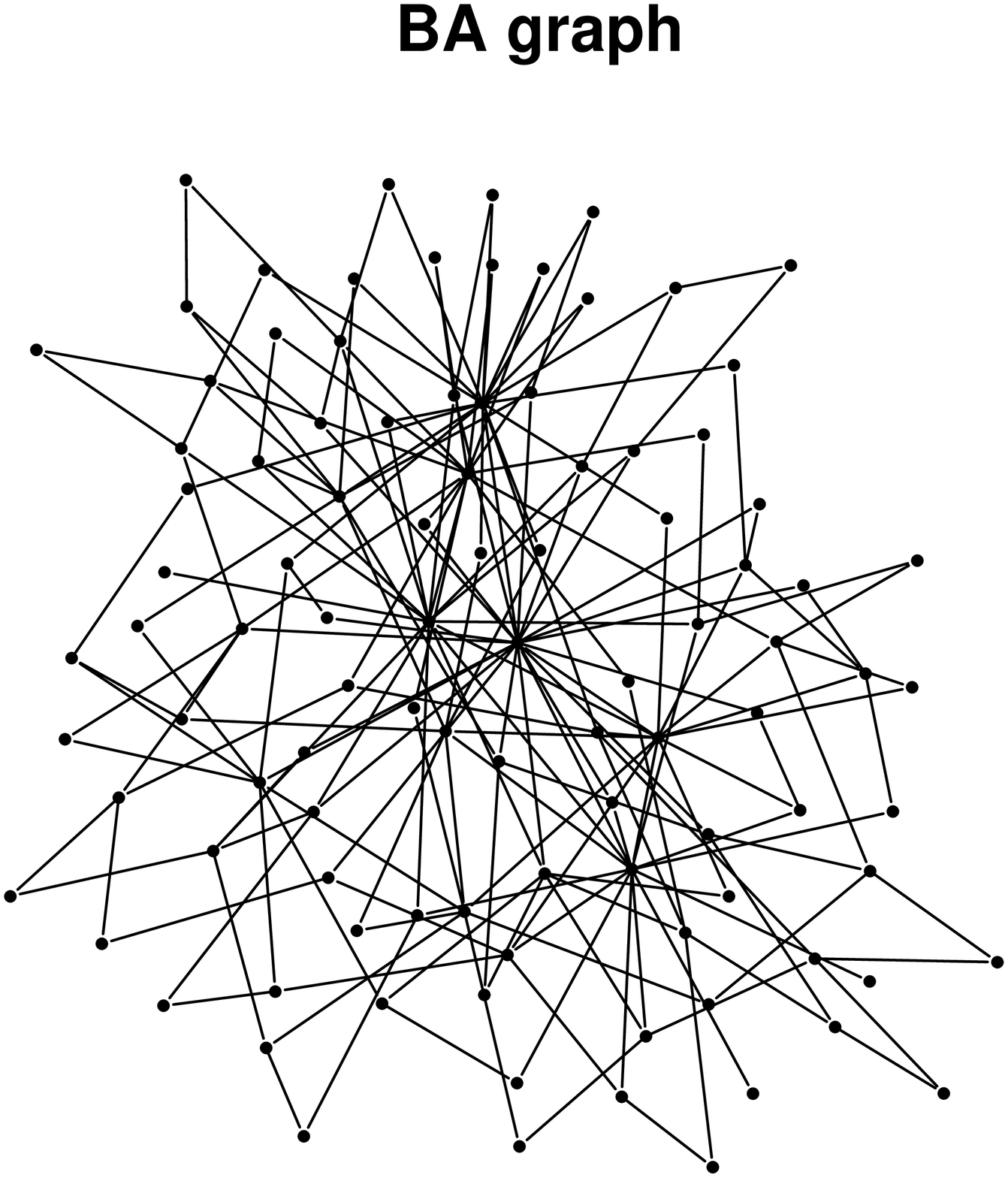}
     \includegraphics[scale = 0.2,clip = true ]{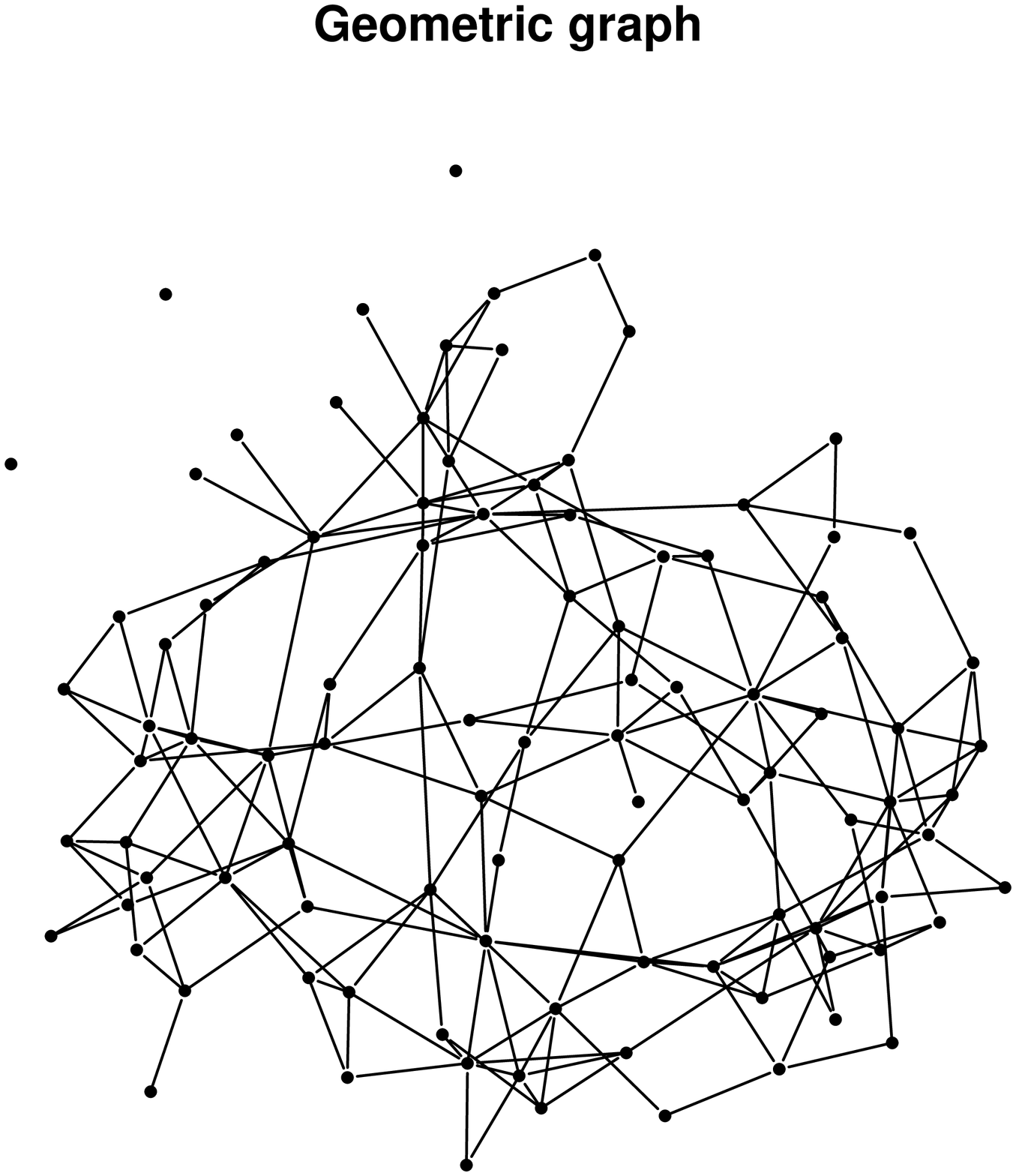}
     \end{center}
     \vspace{-1cm}
     \caption{Plots of ER, BA, and geometric random graphs of size $p=100$
        and average degree $\bar{d}=4$. \label{fig:graphs}}
 \end{figure}

Figure~\ref{fig:graphs} shows examples of three network topologies.
The first is an Erd\"{o}s-R\'{e}nyi (ER) random graph\cite{Erdos}, a
classical form of random graph in which vertex pairs $i,j$ are
assigned edges according to independently and identically
distributed Bernoulli random variables (i.e., coin flips). The
degree distribution of an ER network is concentrated around its mean
and has tails that decay exponentially fast. The second is a random
graph generated according to the Barab\'{a}si and Albert
model~\cite{BA}, which was originally motivated by observed
structure in the World Wide Web.  The defining characteristic of the
BA model is that the derived network has a degree distribution of a
power-law form, with tails decreasing like $d^{-3}$ for large $d$.
Therefore, the BA networks tend to contain many vertices with only a
few neighbors, and a few vertices with many neighbors.  Lastly, we
also use a geometric random graph model, such as might be
appropriate for modeling spatial networks.
Following\cite{Meinshausen}, vertices in the graph are uniformly
distributed throughout the unit square $[0,1]^2$, and each vertex
pair $i,j$ has an edge with probability $\phi\left(w_{ij}
p^{1/2}\right)$, where $\phi(\cdot)$ is the standard normal density
function and $w_{ij}$ is the Euclidean distance in $[0,1]^2$ between
$i$ and $j$. In all three cases, the random graph was of size
$p=100$ and had average degree $\bar{d}=4$.

\begin{figure}
     \begin{center}
     \includegraphics[scale = 0.32,clip = true]{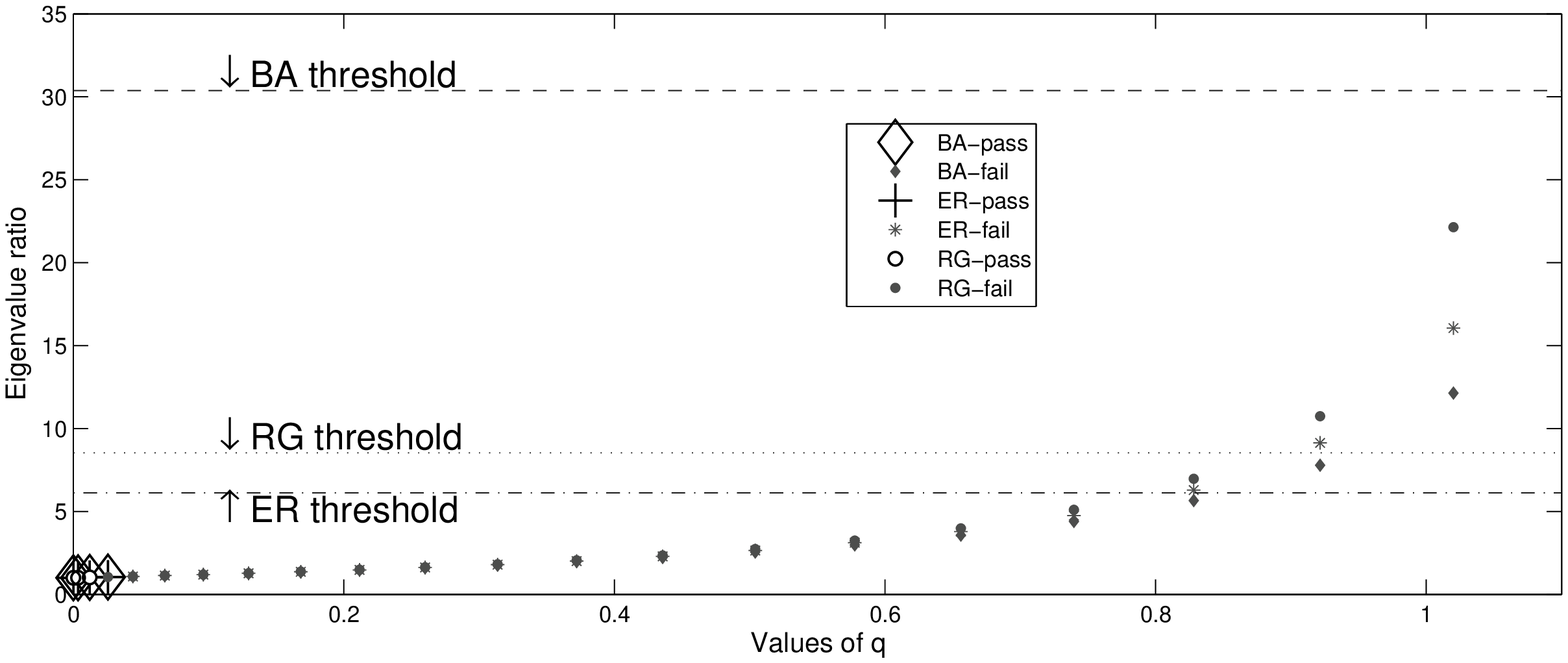}
     \end{center}
     \vspace{-2mm}
     \caption{Plots of eigenvalue ratio for ER, BA, and geometry graph under
     different values of $q$.}
     \label{fig:qtest}
 \end{figure}

In Figure~\ref{fig:qtest} we show the eigenvalue ratio in
(\ref{cond3}), under the simplified covariance structure in
(\ref{eq:toy.cov}), for these ER, BA,and geometric random graphs, as
a function of $q$.
The horizontal lines represent the theoretical eigenvalue ratio
bound given by Theorem 1. The open symbols (including the `plus'
symbol) indicate graphs that satisfy the condition in Theorem 2,
while the filled symbols indicate graphs that do not satisfy the
condition. We can see from the plot that the condition in Theorem~2
clearly is conservative, since as a function of $q$ it ceases to
hold long before the inequality in (\ref{cond3}) is violated.


\subsection{ The sparsity constraint}

The second condition in Theorem~1, given in (\ref{cond4}), can be
read as a condition on the sparsity $r=S/(p-1)$ of the precision
matrix $\Omega\propto I-B$, and therefore a condition on the
sparsity of our network graph $G$. The analytical form of the
function $\rho(\cdot)$ is
\begin{equation}
\rho(r) = (1+f(4r))^2+2(1+f(5r))^2-3 \enskip , \label{eq:def.rho}
\end{equation}
where $f(r) = \sqrt{p/n}\left( \sqrt{r} + \sqrt{2H(r)}\right),$ and
$H(r) = -r\log(r) -(1-r)\log(1-r)$ is the entropy function. While it
is not feasible to produce a closed-form solution in $r$ to the
inequality (\ref{cond4}), it is straightforward to explore the space
of solutions numerically.

\begin{figure}[h]
\begin{center}
\includegraphics[scale=0.5]{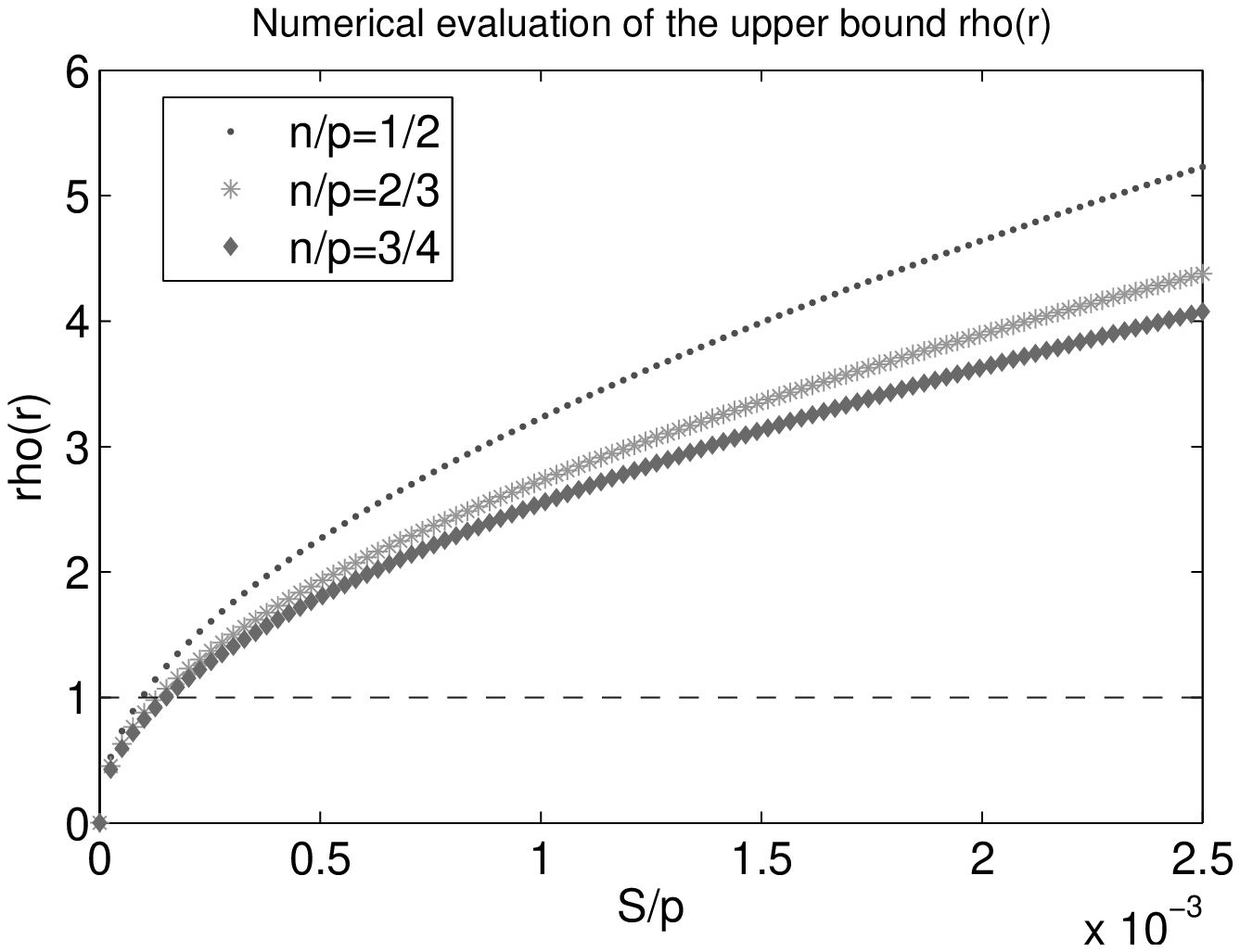}
\includegraphics[scale=0.5]{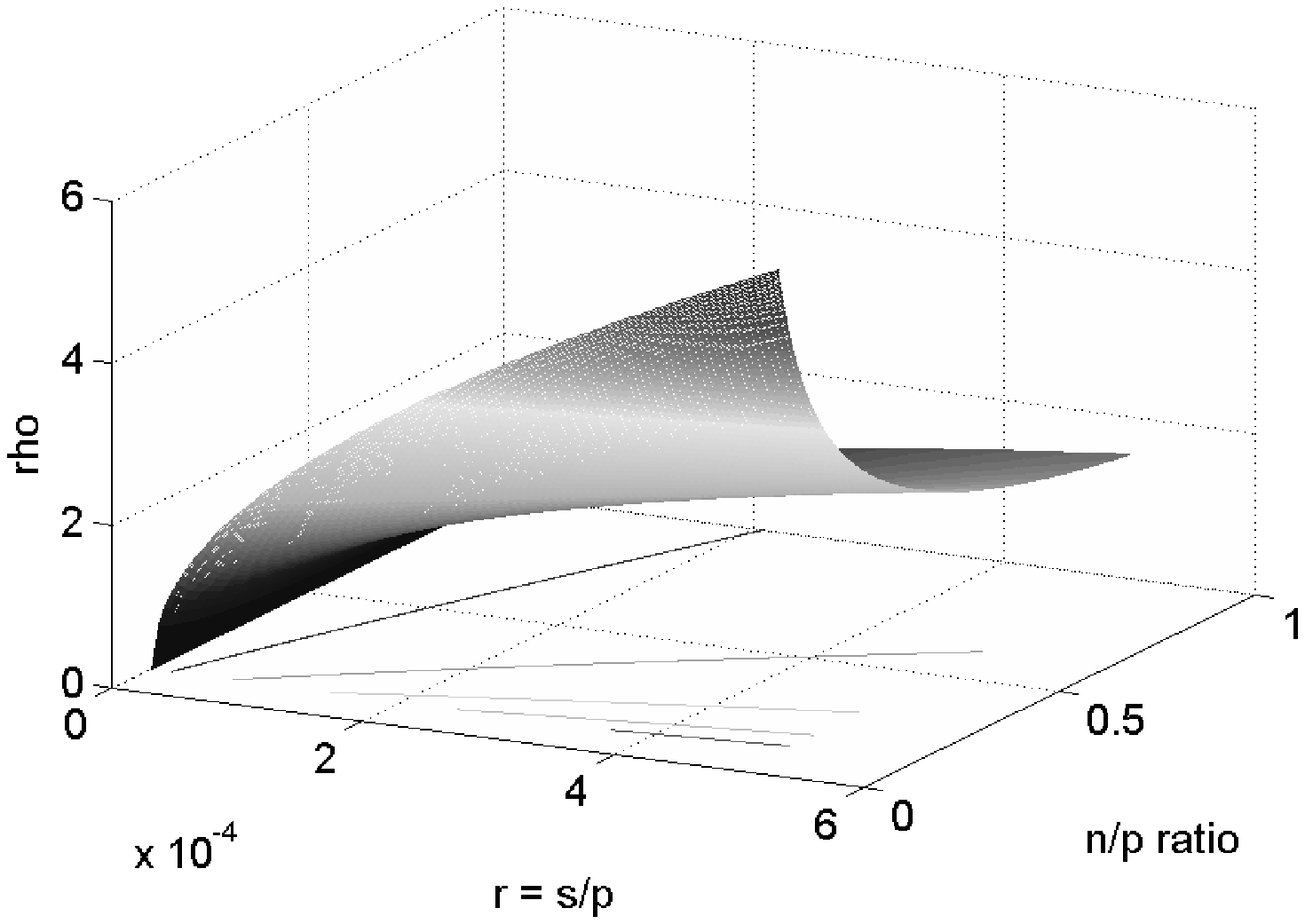}
\caption{2D plot and 3D plot showing the behavior of $\rho_{n/p}(r)$
for three values of the ratio $n/p$.  \label{fig:rho.plots}}
\end{center}
\end{figure}

Note that $\rho(r)$ actually is a function of the three parameters
$S$, $p$, and $n$ through the two ratios $S/(p-1)$ and $n/p$.  In
practice we expect both ratios to be in the interval $(0,1)$.  Shown
in Figure~\ref{fig:rho.plots} is $\rho(\cdot)$, as a function of
$r$, for a handful of representative choices of $n/p$.  We see from
the plot that the theory suggests, through condition (\ref{cond4}),
that the sparsity $r$ should be bounded by roughly $1\times
10^{-4}$. Our numerical results, however, shown later, indicate that
the theory is quite conservative, in that, for example, for our
simulations we successfully used networks with sparsity on the order
of $r=0.04$. Analogous observations have been made in~\cite{Candes}.
Also shown in Figure~\ref{fig:rho.plots} is a 3D plot of
$\rho(\cdot)$, as a function of both $r$ and $n/p$. In this plot,
the dark area corresponding to the innermost contour line satisfies
the condition that $\rho(r)<1$. Again, the value of the information
shown here is primarily as an indication of the existence of
feasible combinations of $S$, $p$, and $n$ allowing for the accurate
estimation of the rows of $B$.


\section{Accuracy of the network filtering}
\label{sec:est.phi}

With the accuracy of $\hat{B}$ quantified, we turn our attention to
the effectiveness of our filtering of the network effects.
Specifically, in the following theorem we characterize the behavior
of $\hat\phi$, defined in (\ref{eq:phi.hat}), as a predictor of
$\hat\phi^{ideal}$, defined in (\ref{eq:phi.hat.ideal}).


\begin{Theorem}
{\it Suppose $\tilde Y$ is a $p\times 1$ vector of test data,
obtained according to the model defined in (\ref{test}) and
(\ref{testdist}).  Let $\hat\phi = (I-\hat B)\tilde Y$ be defined as
in (\ref{eq:phi.hat}) and let $\Delta=B-\hat B$. Then conditional on
$\hat B$, $\hat\phi$ has a multivariate normal distribution, with
expectation and variance \small{
\begin{equation}
E\left[\hat{\phi}\,\big\vert\,\hat{B}\right] =\phi +
\Delta(I-B)^{-1}\phi
    \label{eq:cond.phihat.mean}
\end{equation}}
\small{
\begin{equation}
\Var\left[\hat{\phi}\,\big\vert\,\hat{B}\right] = (I-B)\sigma^2 +
    \left[\,\Delta(I-B)^{-1} \Delta^T + 2\Delta\,\right]\,\sigma^2
    \label{eq:cond.phihat.var} \enskip .
\end{equation}}
Furthermore, under the conditions of Theorem~1, element-wise we have
\small{
\begin{equation}
\left\vert\, E\left[(\hat{\phi}_i - \phi_i) \,\big\vert\, \hat
B\right]
    \,\right\vert
\le ||\phi||_2\, \left[\, (C\sigma^2\zeta_n^+)^{1/2}\,
    \lambda_{max}\left((I-B)^{-1}\right)\,\right]
\label{eq:phihat.bias.bnd}
\end{equation}}
and
\begin{equation}
\Var\left[\hat{\phi}_i\,\big\vert\, \hat B\right] \le
\sigma^2\left[\, 1 + C\sigma^2\zeta_n^+
        \lambda_{max}\left((I-B)^{-1}\right)\,\right] \enskip ,
\label{eq:phihat.var.bnd}
\end{equation}
with overwhelming probability, where $C>0$ and $\zeta_n^+$ are as in
Theorem~1. }
\end{Theorem}
\medskip

Proof of this theorem may be found in Appendix~C.  Recall that
$\hat\phi^{ideal}$ in (\ref{eq:phi.hat.ideal}) is distributed as a
multivariate normal random vector, with expectation $\phi$ and
variance $(I-B)\sigma^2$.  Equations (\ref{eq:cond.phihat.mean}) and
(\ref{eq:cond.phihat.var}) show that our predictor $\hat\phi$ mimics
$\hat\phi^{ideal}$ well to the extent that our error in estimating
$B$ -- that is, those terms involving $\Delta$ -- are small.
Theorem~1 quantifies the magnitude of the rows $\Delta_{i\bullet} =
B_{i\bullet} - \hat B_{i\bullet}$ of $\Delta$, from which we obtain
the term $C\sigma^2\zeta_n^+$ in our bounds on the element-wise
predictive bias in (\ref{eq:phihat.bias.bnd}) and variance in
(\ref{eq:phihat.var.bnd}).

\noindent {\em Remark 4:} In the case that there are no external
effects exerted upon our system i.e., $\phi=0$, the elements
$\hat\phi^{ideal}_i$ of the ideal estimate $\hat\phi^{ideal}$ are
just identically distributed $N(0,\sigma^2)$ noise.  This case
corresponds to the intuitive null distribution we might use to
formulate our detection problem as a statistical hypothesis testing
problem.  The implication of the theorem is that, in using
$\hat\phi$ rather than $\hat\phi^{ideal}$, following substitution of
$\hat B$ for $B$, the price we pay is that the elements $\hat\phi_i$
are instead distributed as $N(0,\tilde\sigma^2_i)$, where the
$\tilde\sigma^2_i$ differ from $\sigma^2$ by no more than
$C\sigma^2\zeta_n^+\lambda_{max}\left((I-B)^{-1}\right)$.  Treating
$\lambda_{max}\left((I-B)^{-1}\right)$ as a constant for the moment,
this term is dominated by $C\sigma^2\zeta_n^+$ i.e., our error in
estimating the rows of $B$. Hence, for example, if $\sigma^2 =
O(n^{-\omega})$ with $\omega>1$, as in Remark~1, then the variances
$\tilde\sigma^2_i$ will also be $O(n^{-\omega})$.
\smallskip

\noindent {\em Remark~5:}  Suppose instead that $\phi =
(0,\ldots,0,\phi^*,0,\ldots,0)'$, for some $\phi^*>0$. This case
corresponds to the simplest alternative hypothesis we might use,
involving a non-trivial perturbation, and is a reasonable proxy for
the type of `genetic perturbations' (e.g., from gene knock-out
experiments) considered in Cosgrove {\it et al.}\cite{Elissa}. Now
the bias is potentially non-zero, even for units $i$ with
$\phi_i=0$.  But, again treating
$\lambda_{max}\left((I-B)^{-1}\right)$ as a constant, and assuming
$\sigma^2 = O(n^{-\omega})$, this bias will be only negligibly worse
than the $O(n^{-(\omega-1)/2})$ magnitude of the ideal standard
deviation $\sigma$. And the variance will again be $O(n^{-\omega})$.
Therefore, we should be able to detect single-unit perturbations
well for $\phi^*$ sufficiently above the noise.  Our simulation
results in the next section confirm this expectation.
\smallskip

Now consider the term $\lambda_{max}\left((I-B)^{-1}\right)$, which
reflects the effect of the topology of $G$ on our ability to do
detection with network filtering.  This term will not necessarily be
a constant in $n$, due to the role of $n$ in the bounds
(\ref{cond3}) and (\ref{cond4}) of Theorem~1, constraining the
behavior of $\Sigma = (I-B)^{-1}\sigma^2$. The following lemma lends
some insight into the behavior of this term in the case where the
precision matrix $\Omega=\Sigma^{-1}$ again has the simple form
specified in (\ref{eq:toy.cov}).  The proof may be found in
Appendix~C.
\medskip

\begin{lemma}
{\it Suppose that $\Sigma^{-1} = I + q D^{-1/2}\, A\, D^{-1/2}$, as
in (\ref{eq:toy.cov}).  Then
\begin{equation}
\lambda_{max}\left((I-B)^{-1}\right)\le
\frac{\sqrt{d_{max}}}{q+\sqrt{d_{max}}}\,
        \left(\frac{1 + \sqrt{d_{max}/n}}
            {1 - \sqrt{d_{max}/n}}\right)^2 \enskip .
\label{eq:lmin.bnd}
\end{equation}
}
\end{lemma}
\medskip

\noindent {\em Remark~6:} Because we assume that the network $G$
will be sparse, and that $d_{max} < n$, the above result indicates
that the term $\lambda_{max}\left((I-B)^{-1}\right)$ can be treated
under our simplified covariance as a constant essentially with
respect to $\sigma^2\zeta_n^+$ in expressions like
(\ref{eq:phihat.bias.bnd}) and (\ref{eq:phihat.var.bnd}).


\section{Simulation Results}
\label{sec:simul}

\subsection{Background}

In this section, we use simulated network data to further study the
performance of our proposed network filtering method.  The data are
drawn from the models for training and test data defined in
Section~II, with randomly generated covariance matrices $\Sigma$. We
define these covariances through their corresponding precision
matrices $\Omega=\Sigma^{-1}$, which are obtained in turn by (i)
generating a random network topology $G=(V,E)$, and then (ii)
assigning random weights to entries in $\Omega$ corresponding to
pairs $i,j$ with edges $\{i,j\}\in E$.  These collections of weights
are then rescaled in a final step to coerce $\Omega$ into the form
$I-B$ and, if necessary, to enforce positive definiteness.  For the
topology $G$, we use the three classes of random network topologies
$G=(V,E)$ described above in Section~III.A i.e., the ER, BA, and
geometric networks. For each choice of network, we use $p=100$
nodes, each of which has an average degree of $\bar{d}=4$.  The
adjacency matrices $A$ of the ER and BA model are generated randomly
using the algorithms listed in \cite{Batageljy}, while that of the
geometric network is generated according to the method described
in~\cite{Meinshausen}.

In implementing our network filtering method, the Lars~\cite{lars}
implementation of the Lasso optimization in (\ref{eq:lasso.est}) was
used, on training data sets of various sample sizes for each
network. The Lasso regularization parameter $\mu$ was chosen by
cross-validation. To generate testing data, we used single-unit
perturbations of the form $\phi = (0,\ldots,0,\phi^*,0,\ldots,0)'$,
where $\phi^*>0$ is in the $i$-th position, for each
$i=1,\ldots,100$.  Since $\sigma^2$ in our simulation is effectively
set to $1$, $\phi^*$ can be interpreted as the signal-to-noise ratio
(SNR) of the underlying perturbation.  In our simulations, we let
$\phi^*$ range over integers from $1$ to $20$.  Our final objective
of detection is to find the position of the unit at which the
external perturbation occurred.  In our proposed network filtering
method, we declare the perturbed unit to be that corresponding to
the entry of $\hat\phi$ with largest magnitude i.e., $\hat i = \arg
\max_{1\le i\le p} |\hat\phi_i|$.

In each experiment described below, our method is compared with two
other methods.  The first, called `True', is that in which the ideal
$\hat\phi^{ideal}$ is used instead of $\hat\phi$, which presumes
knowledge of the true $B$. The second, called `Direct', is that in
which the actual testing data $\tilde Y$ i.e., the data without
network filtering, are used instead of $\hat\phi$.  In both cases,
we declare the perturbed unit to be that corresponding to the entry
of largest magnitude.  The `True' method gives us a benchmark for
the detection error under the ideal situation that we already have
all of the network information, while the `Direct' method is a
natural approach in the face of having no information on the
network. By comparing our method with the two, we may gauge how much
is gained by using the network filtering method.  In all cases,
performance error is quantified as the fraction of times a perturbed
unit is not correctly identified i.e., the proportion of
mis-detections. Results reported below for all three methods are
based in each case upon $30$ replicates of the testing data.  Our
plots show average proportions of mis-detections and one standard
deviation.

\subsection{Results}

First we present the results from an experiment where $\Sigma$ is
defined according to the simple formulation given in
(\ref{eq:toy.cov}), the definition that underlays the results in
Theorem~2 and Lemma~1.  That is, we define $\Sigma$ in terms of just
the (random) adjacency structure of our three underlying networks,
scaled by an appropriate choice of $q$ to ensure positive
definiteness.  We may think of this case, from the perspective of
the simulation design described above, as one with a particular
non-random choice of weights for edges in the network $G$ i.e.,
where $B= - q D^{-1/2} A D^{-1/2}$.

\begin{figure}[h]
\begin{center}
\includegraphics[scale = 0.33]{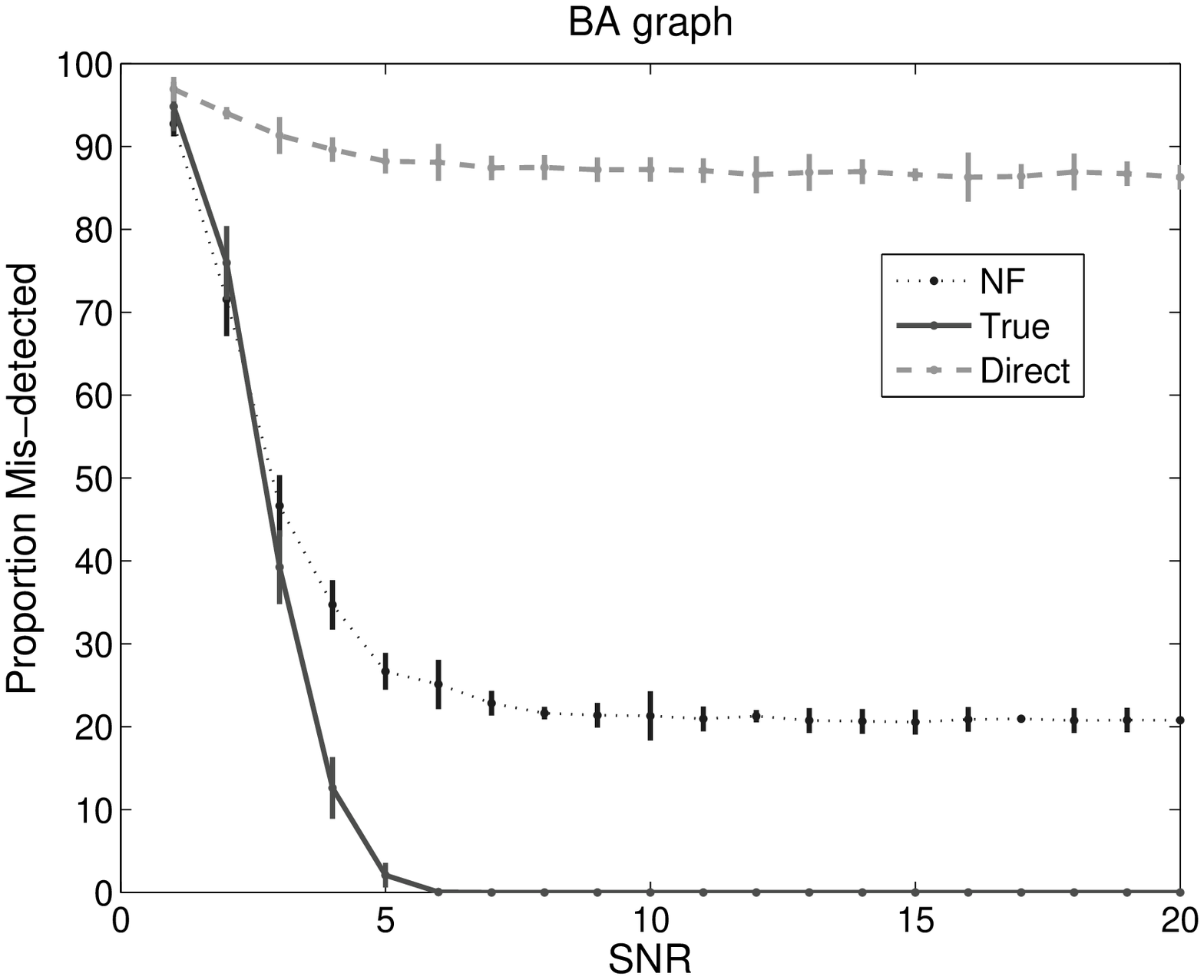}
\includegraphics[scale = 0.33]{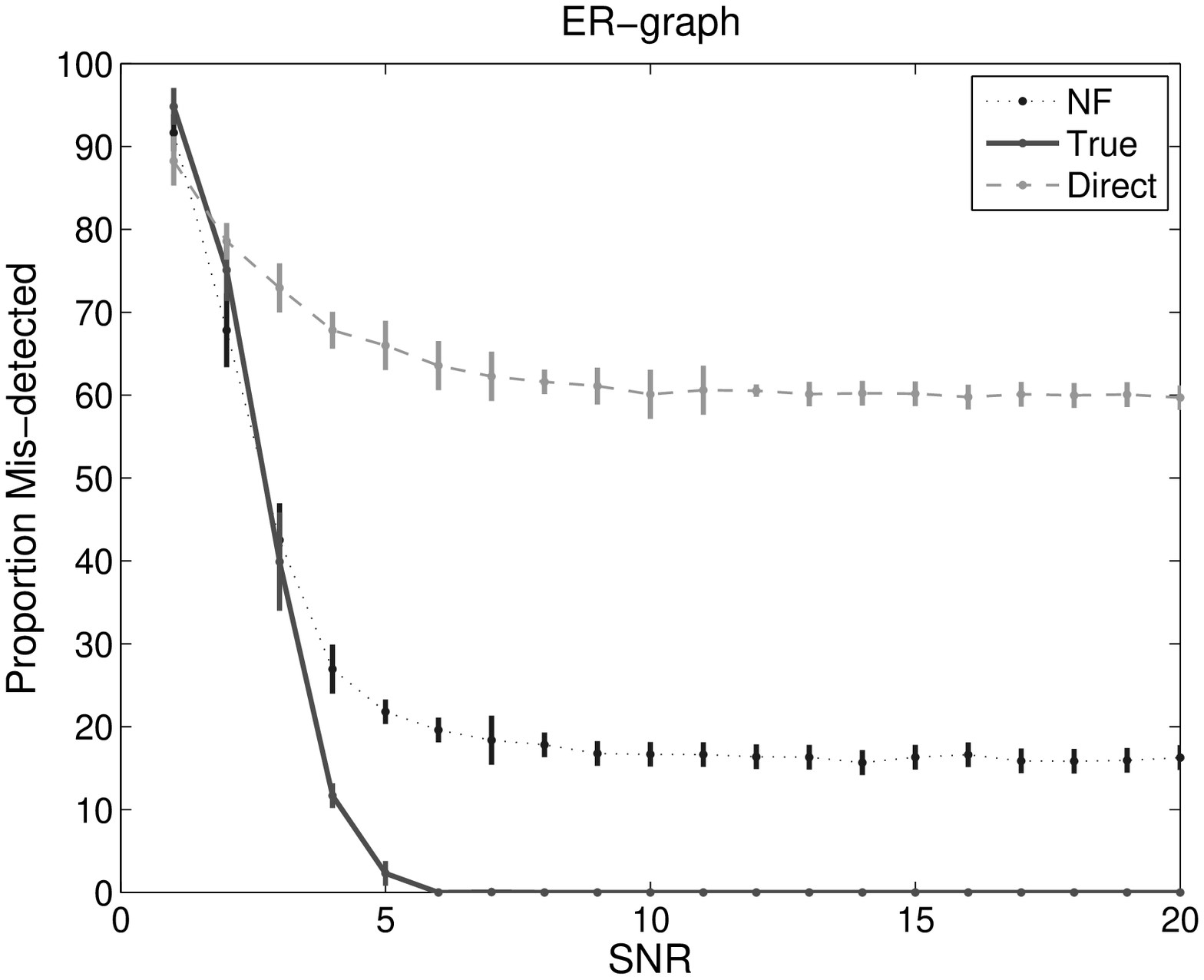}
\includegraphics[scale = 0.33]{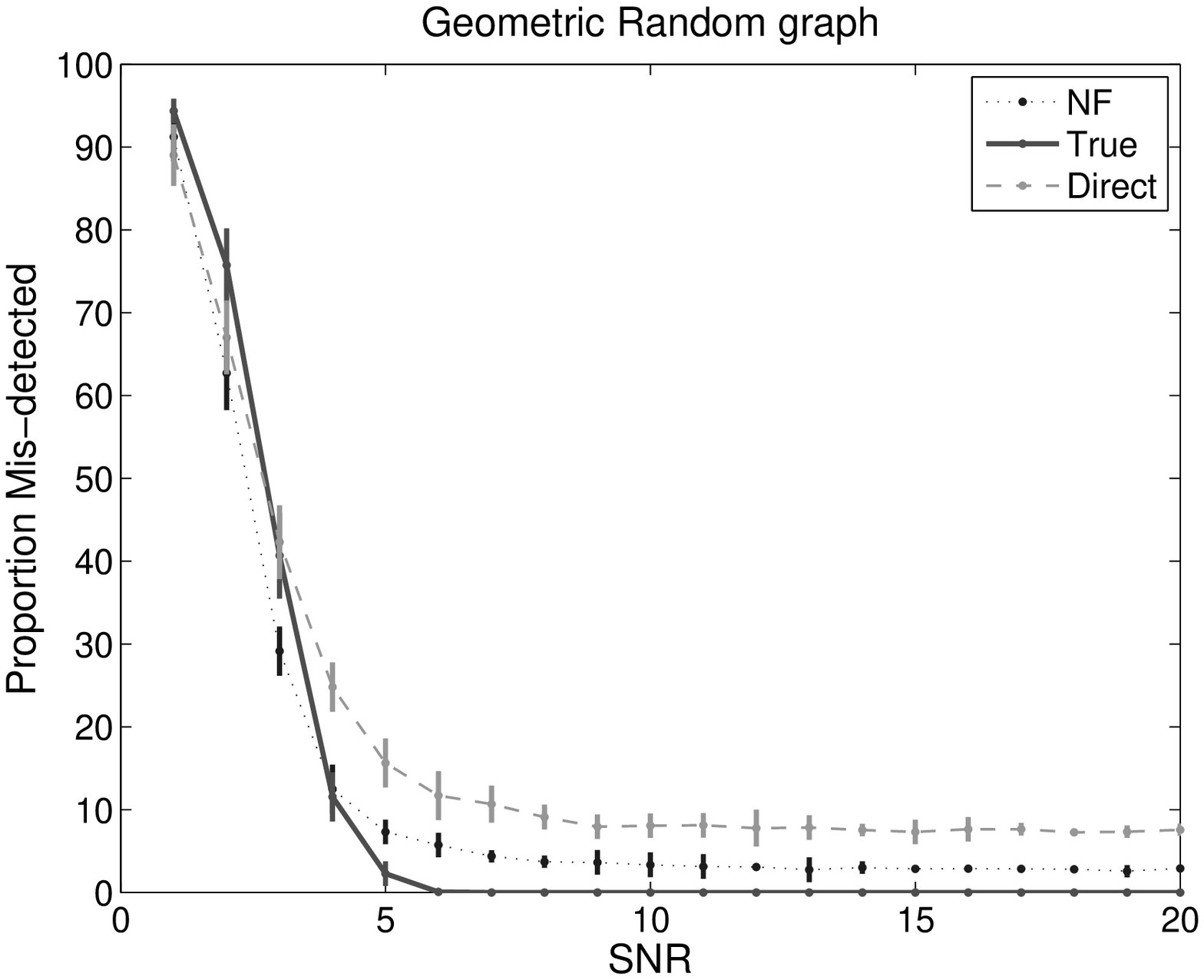}
\end{center}
\caption{Plots of the proportion of mis-detections versus
signal-to-noise ratio, for the BA, ER and geometric random networks,
based on the simplified covariance model in (\ref{eq:toy.cov}),
using $q= 1.25$, $p=100$ and $n=50$. Error bars indicate one
standard error over $30$ test datasets.  \label{same}}
\end{figure}

Figure~\ref{same} shows the average proportions of mis-detections,
as a function the SNR, for these three models.  Note that since the
underlying graphs are random, there is some variability in such
detection results from simulation to simulation. However, these
plots and the others below like them are representative in our
experience. From the plots in the figure, we can see that in all
cases the network filtering offers a significant improvement over
the `Direct' method, and in fact comes reasonably close to matching
the performance of the `True' method, with mis-detections at a rate
of roughly 5-25\% for high SNR. Performance differs somewhat with
respect to networks of different topology. The network filtering
method shows the most gain over the `Direct' method with the BA
network. This phenomenon is consistent with our intuition: the
distribution of edges in the BA network is the least uniform one,
and certain choices of perturbed unit (i.e., perturbed units $i$
with large degree $d_i$) will enable the effects of perturbation to
spread comparatively widely. Hence obtaining and correcting for the
internal interactions among units in the network is particularly
helpful in this case.

\begin{figure*}[h]
\begin{center}
\includegraphics[scale = 0.8,clip = true]{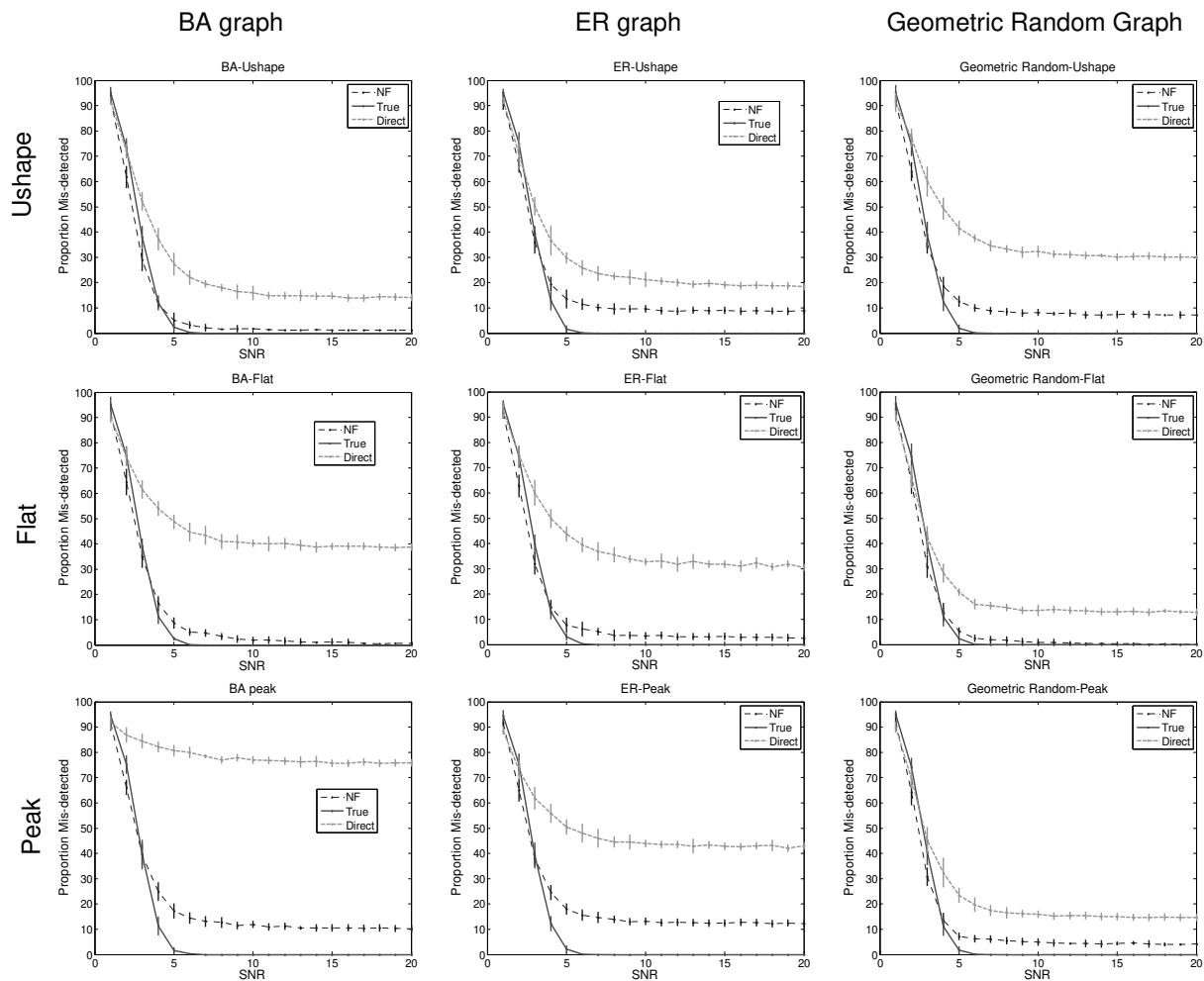}
\end{center}
\vspace{-0.5cm} \caption{Plots of proportions of mis-detections
versus signal-to-noise ratio.  Columns: BA (left), ER (middle), and
geometric (right) random networks.  Rows: U-shaped (top), flat
(middle), and peaked (bottom) choice of weight distributions,
generated according to $Beta(1/2,1/2)$, $Beta(1,1)$ and $Beta(2,2)$
distributions, respectively. \label{fig:beta.results}}
\end{figure*}

Now consider the assignment of random weights to edges in $G$, which
allows us to generate a richer variety of models.  For this purpose,
we choose the family of beta distributions $Beta(a,b)$ from which to
draw weights $B_{ij}$ independently for each edge $\{i,j\}\in E$.
Three different classes of distributions were used i.e.,
$Beta(1,1)$, $Beta(1/2,1/2)$, and $Beta(2,2)$, which gives flat
(uniform), U-shape, and peaked shape forms.  Shown in
Figure~\ref{fig:beta.results} are the results of our network
filtering method, the `True' method, and the `Direct' method, for
each of these three choices of weight distributions, for each of the
three network topologies. The same (random) network topology is used
in each plot for each type of network.

Broadly speaking, these plots show that the performance of network
filtering in the context of randomly generated edge weights
$B_{ij}$, as compared to that of the `True' and `Direct' methods, is
essentially consistent with the case of fixed edge-weights
underlying the plots in Figure~\ref{same}.  However, there are some
interesting nuances. For example, in the case of `Flat' weights,
network filtering in fact is able to match the performance of `True'
for all three classes of graphs.  On the other hand, in the ER
random network topology this matching occurs only when the
edge-weight distribution is flat (i.e., $Beta(1,1)$), and in the BA
random network topology, when the distribution is either $U$-shaped
(i.e., $Beta(1/2,1/2)$) or flat (i.e., $Beta(1,1)$).  Nevertheless,
the qualitatively similar performance across choice of edge-weight
distribution suggests that most important element here is the
network structure, indicating connection between pairs of units,
with the strength of connection being secondary.

\begin{figure}[h]
\begin{center}
\includegraphics[width = 5.8cm,clip = true]{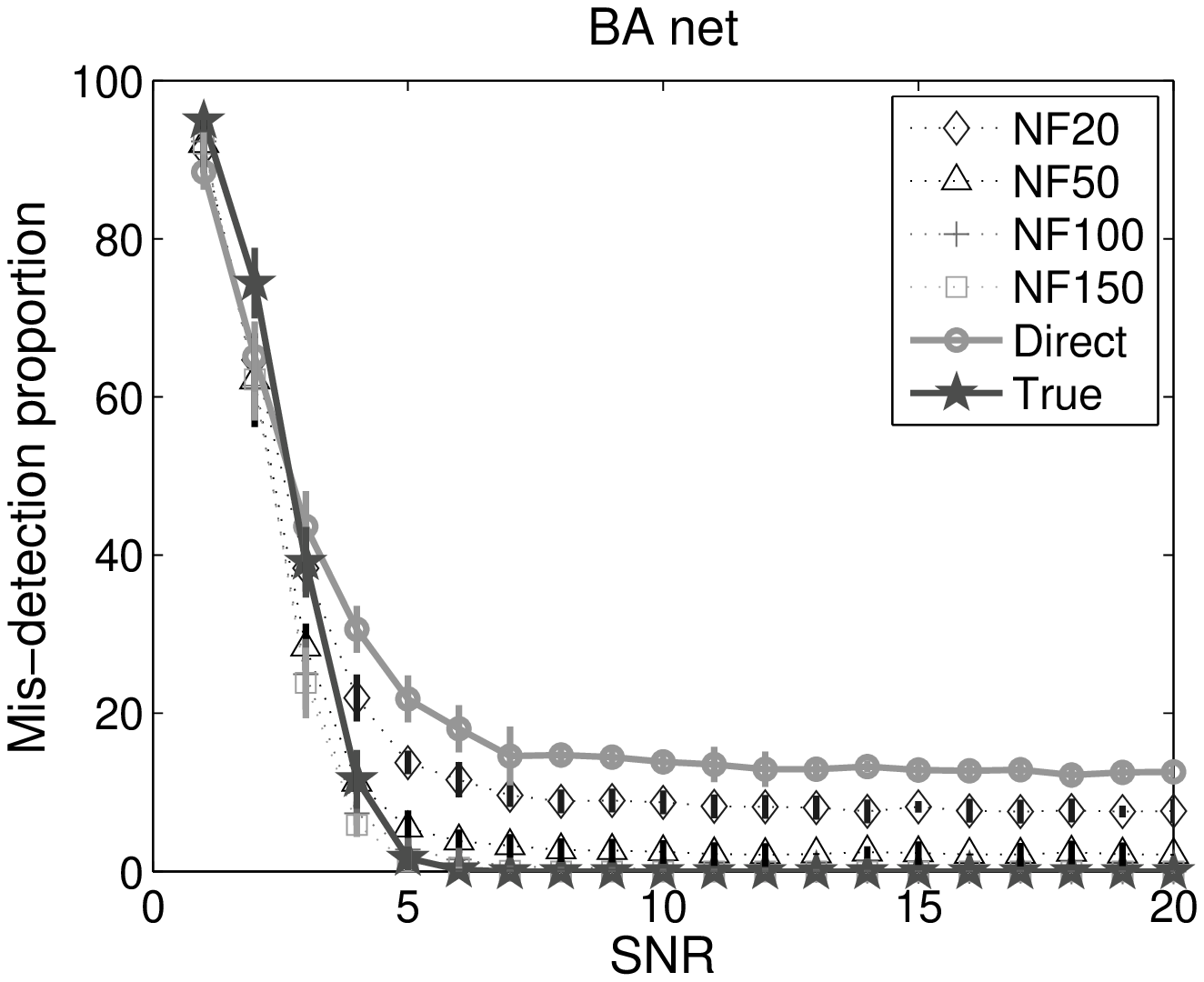}
\includegraphics[width = 5.8cm,clip = true]{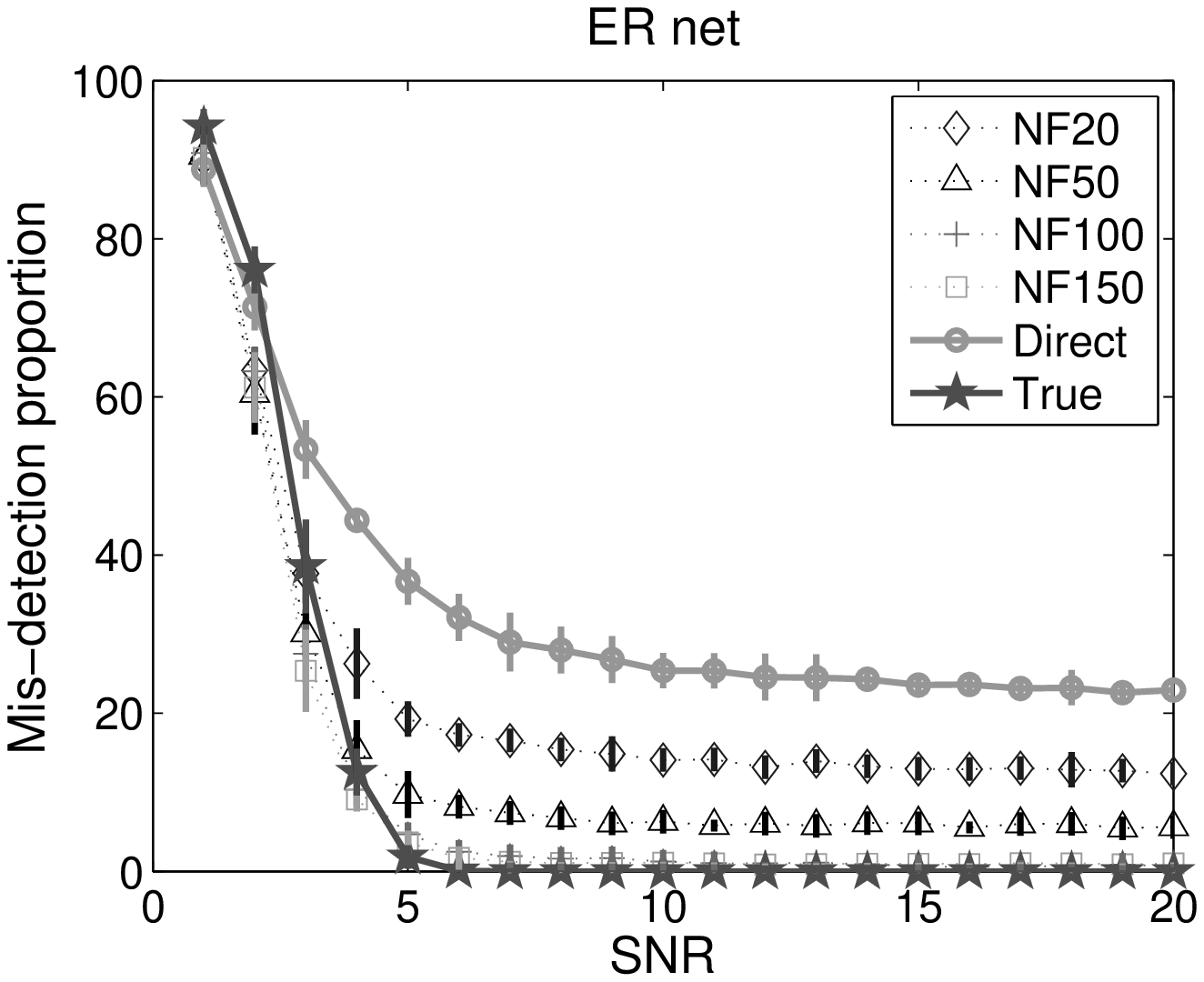}
\includegraphics[width = 5.8cm,clip = true]{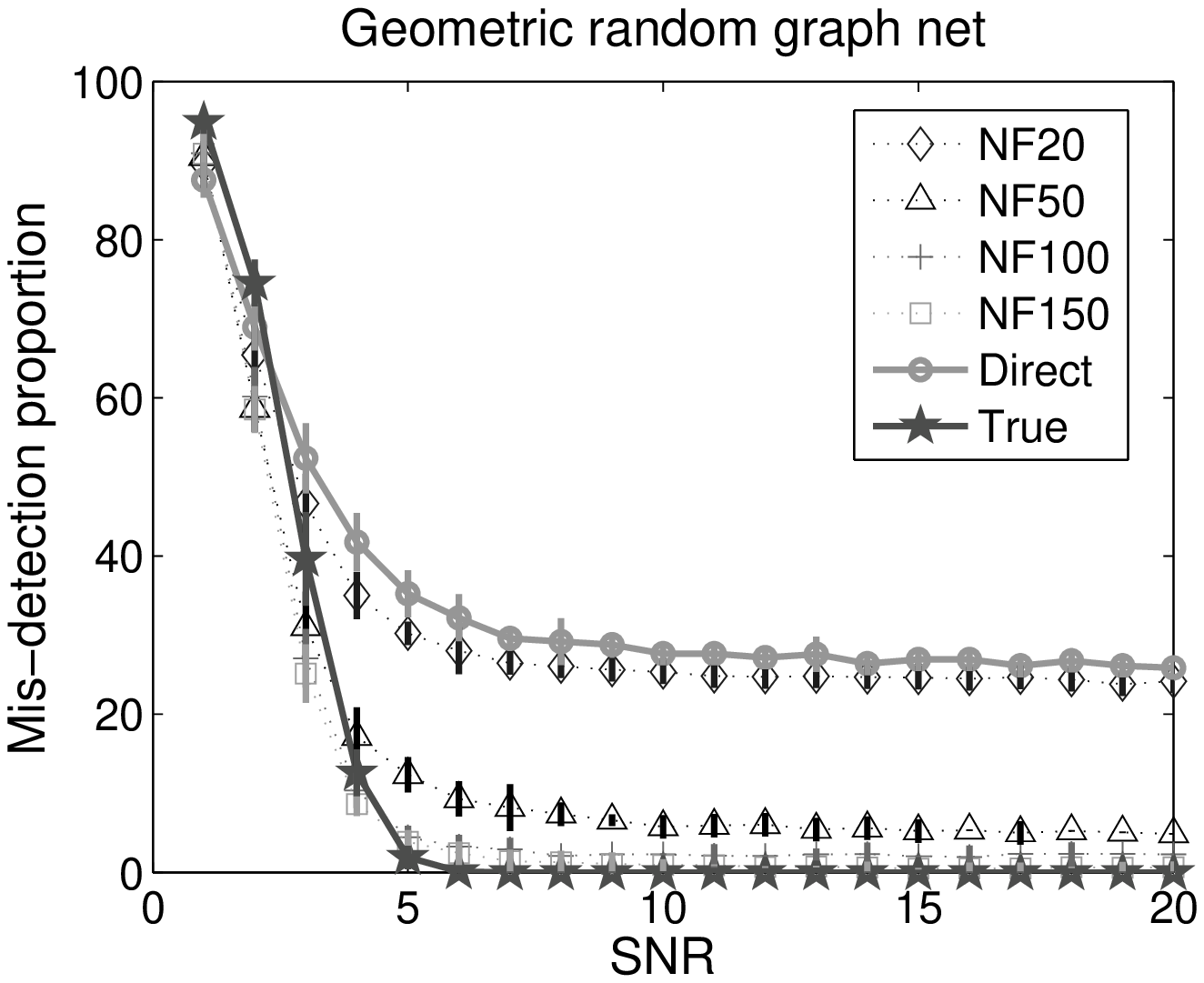}\\
\end{center}
\vspace{-0.5cm} \caption{Plots of proportion of mis-detections
versus signal-to-noise ratio for the BA (left), ER (middle), and
geometric (right) random network models, for sample sizes
$n=20,50,100$, and $150$.  \label{fig:nvaries}}
\end{figure}

Finally, we consider the effect of sample size $n$ and, therefore
implicitly, the extent to which the condition in (\ref{cond3}) on
the structure of the covariance matrix $\Sigma$ may be relaxed.  For
the same networks used in the simulations described above, with
$p=100$ units, we varied the sample size $n$ to range over $20, 50,
100$, and $150$.  Weights of the network edges are set according to
a $Beta(2,2)$ distribution, which is the `peak' case. Training and
testing data were generated as before. The results of using network
filtering in these different settings are shown in
Figure~\ref{fig:nvaries}.  Again, our network filtering method is
seen to work similar to above.  Even for a sample size as small as
$n=20$, our method still does better than the `Direct' method in all
three models, particularly under the BA and ER models\footnote{We
note that some care must be used in fitting Lasso with $n=p$, due to
numerical instabilities that can arise.  This issue affects any
method attempting to estimate the inverse of a covariance matrix (as
is implicitly being done here).  Kr\"amer~\cite{kramer} describes
how a re-parameterization of the Lasso penalty can be used to avoid
this problem.}.

On a final note, we point out that in all of the experiments the
richness of network models studied is much broader than suggested by
our theory.  As was mentioned earlier, the concentration
inequalities we use can be expected to be conservative in nature,
and therefore some of the bounds are more restrictive than practice
seems to indicate is necessary. For example, in our simulations
involving the geometric random graph
in Figure~\ref{fig:graphs}, with a sample size of $n=50$ and $S = 9$, 
the theoretical bound (\ref{cond3}) for the eigenvalue ratio is
$6.12$, while the actual value achieved by this ratio is $219$ in
this instance. Also, the maximum degrees of the graphs in most of
the simulations are larger than the average degree 4, and hence the
sparseness rate $S/p > 4/100$, which is already larger than those
theoretical sparse rates suggested by Figure~1. Yet still we
observed the network filtering method to perform quite well.
It is an interesting open question to see if the theory can be
extended to produce bounds like (\ref{cond3}) that more accurately
reflect practice, to serve as a better practical guides for users.

\section{Discussion}
\label{sec:disc}

The concept of network filtering considered in this paper was first
proposed by Cosgrove {\it et al.}~\cite{Elissa}, as a methodology
for filtering out the effects of `typical' gene regulatory patterns
in DNA microarray expression data, so as to enhance the potential
signal from genetic targets of putative drug compounds. Here we have
formalized the methodology of Cosgrove {\it et al.} and established
basic conditions under which it may be expected to perform well.
Furthermore, we have explored the implications of these conditions
on the topology of the network underlying the data (i.e., a Gaussian
concentration graph). Proof of our results rely on principles and
techniques central to the literature on compressed sensing and,
therefore, like other results in that literature, make performance
statements that hold with over-whelming probability.  Numerical
simulation results strongly suggest a high degree of robustness of
the methodology to departure from certain of the basic conditions
stated in our theorems regarding network topology.  Our current work
is now focused on the development of adaptive learning strategies
that intentionally utilize perturbations (i.e., in the form of the
vectors $\phi$) to more efficiently explore network effects (i.e.,
the matrix $B$).


\appendices
\section{Proof of the First Zonklar Equation}
\section{Proof of Theorem 1}

Theorem~1 jointly characterizes the accuracy of $p$ simultaneous
regressions, each based on the model in (\ref{mainReg}) i.e., for
$i=1,\ldots,p$,
\begin{equation}
y_i = \sum_{j\neq i}^p \beta_{ij}y_j + e_i \enskip .
\label{eq:mainReg.again}
\end{equation}
For convenience, we re-express the above model for a single
regression in the generic form
\begin{equation}
R = X\beta + e \enskip . \label{regression}
\end{equation}
Here $X$ is a $n\times (p-1)$ design matrix with rows sampled i.i.d.
from a multivariate normal distribution $N(0, \mathbb{C})$, with
$(p-1)\times (p-1)$ covariance matrix $\mathbb{C}$; $e$ is an
$n\times 1$ error vector, independent of $X$, with i.i.d.
$N(0,\sigma^2)$ elements; and $R$ is the $n\times 1$ response
vector.

We will make use of a result of Zhu~\cite{Zhu}, which requires the
notion of restricted isometry constants.  Following
Zhu\footnote{This definition differs slightly from that in
Cand{\'e}s~\cite{Candes}.  See~\cite{Zhu} for discussion.}, we
define the $S$-restricted isometry constant $\delta_S$ of the matrix
$X$ as the smallest quantity such that
\begin{equation}
(1-\delta_S) \|c\|_2 \leq \|X_Tc\|_2 \leq (1+\delta_S) \|c\|_2
\label{RIC}
\end{equation}
for all index subsets $T\subset \{1,\ldots,p\}$ with $|T|\leq S$.
Zhu's result is then as follows.

\begin{lemma}[Zhu]
{\it If (i) the number of non-zero entries of $\beta$ is no more
than $S$, (ii) the isometry constants $\delta_{4S}$ and
$\delta_{5S}$ obey the inequality $\delta_{4S}+2\delta_{5S}<1$, and
(iii) the Lasso regularization parameter $\mu$ obeys the constraint
$\mu^2\le C_0 \zeta/S$, for $\zeta=||e||^2_2$, then
\begin{equation}
||\beta - \hat\beta||^2_2 \le \tilde C \zeta\enskip ,
\label{eq:zhu.bnd}
\end{equation}
where
\begin{equation}
\hat\beta = \arg\min_{\beta} \|X\beta - R\|^2_2 +
\mu\|\beta\|_1\enskip . \label{eq:zhu.lasso}
\end{equation}
\label{lem:zhu} }
\end{lemma}

Zhu's first condition is assumed in our statement of Theorem~1.
Therefore, to prove Theorem~1 we need to show, under the other
conditions stated in our theorem, that Zhu's second and third
conditions above hold simultaneously for each of our $p$
regressions, with overwhelming probability.  In addition, we need to
show that the right-hand side of (\ref{eq:zhu.bnd}) is bounded above
by the right-hand side of (\ref{eq:cs.bound}).

\subsection{Verification of Lemma~\ref{lem:zhu}, Condition (ii)}

The essence of what is needed for the restricted isometry constants
is contained in the following lemma.

\begin{lemma}
{\it Suppose $\mathbb{C}_T$ is a sub-matrix of the covariance matrix
$\mathbb{C}$ with columns variables corresponding to these in
$\mathbb{C}$ of the indices in set $T$, where $|T| = S$. Denote the
largest and smallest eigenvalues of any such matrix $\mathbb{C}_T$
as $\lambda_{max}(\mathbb{C}_T)$ and $\lambda_{min}(\mathbb{C}_T)$
respectively. Suppose too that
\begin{equation}
\frac{\lambda_{max}(\mathbb{C}_T)}{\lambda_{min}(\mathbb{C}_T)} \leq
\left(\frac{1+\sqrt{S/n}}{1-\sqrt{S/n}} \right )^2
\quad\hbox{and}\quad \rho(r) < 1 \enskip , \label{eq:lem.rip.cond}
\end{equation}
where $\rho(r)$ is defined as in (\ref{eq:def.rho}). Then the
condition $\delta_{4S}(X)+2\delta_{5S}(X)<1$ holds with overwhelming
probability.\label{lem:RIP} }
\end{lemma}
\medskip

The covariance matrix $\mathbb{C}$ corresponding to any single
regression is a sub-matrix of $\Sigma$ in Theorem~1, and hence so is
any sub-sub-matrix $\mathbb{C}_T$.  By the interlacing property of
eigenvalues (e.g., Golub and van Loan~\cite[Thm 8.1.7]{gvl}), which
relates the eigenvalues of a symmetric matrix to those of its
principle sub-matrices, as long as $\Sigma$ satisfies the eigenvalue
constraint (\ref{cond3}), the matrices $\mathbb{C}_T$ will as well.
So it is sufficient to prove Lemma~\ref{lem:RIP}.

\noindent {\it Proof of Lemma~\ref{lem:RIP}:}\quad Let $X_T$ denote
the $n\times S$ sub-matrix of $X$ corresponding to the subset of
indices $T$.  Since the rows of $X$ are independent samples from
$N(0,\mathbb{C})$, the rows of $X_T$ are independent samples from
$N(0,\mathbb{C}_T)$. Let $\sigma_{i}$ be the \emph{i-th} largest
singular value of $\mathbb{C}_T^{1/2}$ and $\hat{\sigma}_{i}$ be the
\emph{i-th} largest singular value of $n^{-1/2}X_T$. The eigenvalue
condition in the lemma reduces to $(\sigma_1/\sigma_S) \leq
\left[1+(S/n)^{1/2}\right]/ \left[1-(S/n)^{1/2}\right]$. Without
loss of generality, therefore, assume that $\sigma_1 \leq
1+(S/n)^{1/2}$ while $\sigma_S \geq 1-(S/n)^{1/2}$.

Note we can express $X_T$ as $X_T = Z\mathbb{C}_T^{1/2}$, where
$Z\sim N(0, I)$. Then $X_T'X_T =
[\mathbb{C}^{1/2}_T]'Z'Z\mathbb{C}^{1/2}_T$ and hence the
eigenvalues of $X_T'X_T$ are the same as those of $Z'Z\mathbb{C}_T$.
Thus we have
\begin{eqnarray}
\lambda_{\max}\left(\frac{X_T'X_T}{n}\right) &\leq&
\lambda_{\max}\left(\frac{Z'Z}{n}\right)\cdot\sigma_1^2\\
\lambda_{\min}\left(\frac{X_T'X_T}{n}\right) &\geq&
\lambda_{\min}\left(\frac{Z'Z}{n}\right)\cdot\sigma_S^2.
\end{eqnarray}
Let $\hat{\sigma}_{i}^*$ denote the \emph{i-th} largest singular
value of $n^{-1/2}Z$. Therefore we
have\footnote{$(\hat{\sigma}_S^*)^2$ equals the smallest eigenvalue
of $Z'Z/n$, which is $\lambda_{min}$ in~\cite{Candes}. Similar for
$(\hat{\sigma}_1^*)^2.$}
\begin{eqnarray}
\hat{\sigma}_1 &\leq& \hat{\sigma}_1^*\cdot\sigma_1\\
\hat{\sigma}_S &\geq& \hat{\sigma}_S^*\cdot\sigma_s
\label{eigenineq}
\end{eqnarray}
Denote by $\hat{\sigma}_{min}(\cdot),\hat{\sigma}_{max}(\cdot)$ the
smallest and largest singular values of their argument. Notice that
for any index set $T^*\subset T$, we have

\small{
$$
\hat{\sigma}_{min}\left(\frac{X_{T}}{\sqrt{n}}\right) \leq
\hat{\sigma}_{min}\left(\frac{X_{T^*}}{\sqrt{n}}\right) \leq
\hat{\sigma}_{max}\left(\frac{X_{T^*}}{\sqrt{n}}\right) \leq
\hat{\sigma}_{max}\left(\frac{X_{T}}{\sqrt{n}}\right).
$$}
Thus we need only to consider the situation where $|T| = S $ and
choose $\delta$ as the smallest constant that satisfies (\ref{RIC})
for any sub-matrix $X_T$ of size $n\times S$. Therefore, we set $
1-\delta \leq \tilde{\sigma}_S \leq \tilde{\sigma}_1 \leq  1+
\delta, $ where $\tilde{\sigma}_1 = \sup_{|T|=S}\hat{\sigma}_1$ and
$\tilde{\sigma}_S = \inf_{|T|=S}\hat{\sigma}_S$. It then follows
that $ \delta \leq \max( 1-\tilde{\sigma}_S, \tilde{\sigma}_1 - 1) .
$

Now, by the large deviation results in \cite{Ledoux,Candes}, for a
standard Gaussian random matrix $Z \sim N(0, I)$, there are two
relevant concentration inequalities:
\begin{eqnarray}
P\left( \hat{\sigma}_1^* > 1 + \sqrt{S/n} + \eta + t\right) &\leq&
e^{-nt^2/2}\\
P\left( \hat{\sigma}_S^* < 1 - \sqrt{S/n} -\eta -t\right) &\leq&
e^{-nt^2/2},
\end{eqnarray}
where
$\eta$ is an $o(1)$ term.

We can then use the above tools and concentration inequalities to
see how $\delta$ behaves under the conditions described in
Lemma~\ref{lem:RIP}. Notice that, for $\varepsilon >0$, we have
\begin{eqnarray}
&&P\left(1+\delta > (1 + (1+\varepsilon)f(r))^2\right) \\
&\leq& P\left(\max( 2-\tilde{\sigma}_{S}\ , \tilde{\sigma}_1\right)
\geq (1 +
(1+\varepsilon)f(r))^2 )\\
&=&P\left( \{ 2 - \tilde{\sigma}_S \geq (1 +
(1+\varepsilon)f(r))^2\} \cup  \right . \\
&&\left  \{ \tilde{\sigma}_1 \geq (1 + (1+\varepsilon)f(r))^2\}\right) \\
&\leq& P\left(\tilde{\sigma}_S \leq
2-(1+(1+\varepsilon)f(r))^2)\right)\\
&&+ P\left(\tilde{\sigma}_1 \geq (1 + (1+\varepsilon)f(r))^2\right)
\label{eq:key.eigen.bnd}
\end{eqnarray}

Denoting $\gamma = S/n$, we have by (\ref{eigenineq}) that
$\hat{\sigma}_S \geq (1- \sqrt{\gamma})\hat{\sigma}_S^*$. Therefore,
for the term with $\tilde{\sigma}_S$ in (\ref{eq:key.eigen.bnd}), we
have by Bonferroni's inequality that
\begin{eqnarray*}
&&P\left(\tilde{\sigma}_S \leq 2-(1+(1+\varepsilon)f(r))^2)\right)\\
&\leq& \sum_{\{|T| = S\}}P\left(\hat{\sigma}_S(X_T) \leq 2-(1+(1+\varepsilon)f(r))^2)\right)\\
&\leq& |\{|T| = S\} |P\left((1-\sqrt{\gamma})\cdot\hat{\sigma}_S^*
\leq
2-(1+(1+\varepsilon)f(r))^2)\right)\\
&\leq&C(p,S)\cdot
\\&&P\left((1-\sqrt{\gamma})\cdot\hat{\sigma}_S^*
\leq 1-2(1+\varepsilon)f(r)
-((1+\varepsilon)f(r))^2\right)\\
&\leq&C(p,S)P\left(\hat{\sigma}_S^* \leq 1-(1+\varepsilon)f(r)
\right) \enskip .
\end{eqnarray*}

As in~\cite{Candes}, we fix $\eta + t  =
(1+\varepsilon)\sqrt{p/n}\sqrt{2H(r)}$, from which it follows that
$(1+\varepsilon)f(r) \simeq \sqrt{\frac{S}{n}} + \eta + t$ and
$C(p,s)e^{-nt^2/2} \leq e^{-pH(r)\varepsilon/2}$. Hence the above
inequality is equivalent to
\begin{eqnarray*}
&&P\left( \tilde{\sigma}_S \leq 2-(1+(1+\varepsilon)f(r))^2)\right )\\
&\leq&C(p,S)P\left(\hat{\sigma}_S^* \leq
(1-\sqrt{S/n}-\eta - t) \right)\\
&\leq& C(p,S)e^{-nt^2/2}    \\
&\leq& e^{-pH(r)\varepsilon/2}
\end{eqnarray*}

For the term with $\tilde{\sigma}_1$ in (\ref{eq:key.eigen.bnd}),
the analogous inequality
$$P( \tilde{\sigma}_1 \geq (1 + (1+\varepsilon)f(r))^2) \le
 e^{-pH(r)\varepsilon/2}$$
may be verified using a similar argument.
Combining these two probability inequalities for $\tilde{\sigma}_1$
and $\tilde{\sigma}_S$, we have that $ P\left(1+\delta > (1 +
(1+\varepsilon)f(r))^2\right) \leq 2\cdot e^{-pH(r)\varepsilon/2}.$
Ignoring the negligible $\varepsilon$ terms, it
follows that when $p,n$ is large enough $\delta < (1+f(r))^2-1$
holds with overwhelming probability. Defining $\rho(r) =
(1+f(4r))^2+2(1+f(5r))^2-3$, we have that $\delta_{4S} +
2\delta_{5S} < \rho(r)$. Imposing the condition $\rho(r)< 1$, we
obtain that $\delta_{4S} + 2\delta_{5S} <1$, and therefore
Lemma~\ref{lem:RIP} is proved.


\subsection{Verification of Lemma~\ref{lem:zhu}, Condition (iii)
        and the right-hand side of (\ref{eq:cs.bound})}

Let $R^{(i)} = X^{(i)}\beta^{(i)} + e^{(i)}$ denote the regression
equation (\ref{regression}) for the $i$-th of the $p$ simultaneous
regressions in (\ref{eq:mainReg.again}). Condition (iii) of
Lemma~\ref{lem:zhu} requires that the regularization parameter $\mu$
be such that $\mu^2\le C_0 \zeta^{(i)}/S$, where
$\zeta^{(i)}=||e^{(i)}||^2_2$. If so, and assuming of course that
conditions (i) and (ii) are satisfied as well, then the inequality
in (\ref{eq:zhu.bnd}) says that $||\hat\beta^{(i)} -
\beta^{(i)}||^2_2\le C^{(i)} \zeta^{(i)}$. We show that the
condition on $\mu$ in the statement of Theorem~1 i.e., that $\mu^2
\le (C_0\sigma^2\zeta_n^-)/S$, guarantees that Condition (iii) holds
for every $i=1,\ldots,p$ with overwhelming probability. In addition,
we show that $C^{(i)}\zeta^{(i)}\le C\sigma^2\zeta_n^+$ with
overwhelming probability, where $\zeta_n^+ =n\left(1+4(\log_2n /
n)^{1/2}\right)$, and $C$ is bounded by $(1-\rho(r))^{-2}$ times a
constant, as claimed in Remark~1.

Notice that if we have $e\sim N(0,\sigma^2 I_{n\times n})$, then
$||e||^2_2$ is distributed as chi-square on $n$ degrees of freedom.
By~\cite[Sec 4.1, Lemma 4]{Laurent}, for $t'>0$,
$$
P\left\{ \|e\|^2_2 - n\sigma^2 \leq - 4\sigma^2\sqrt{nt'}\right\}
\leq \exp(-t') $$ and
$$ P\left\{ \|e\|^2_2 - n\sigma^2 \geq
4\sigma^2\sqrt{nt'}\right\} \leq \exp(-t')\enskip .
$$
Therefore,
\begin{eqnarray*} &&P\left\{\min_i
\frac{\|e^{(i)}\|^2_2}{n\sigma^2}\leq (1 - 4\sqrt{t'/n})\right\}\\
&& \leq \sum_{i=1}^p P\left\{\frac{\|e^{(i)}\|^2_2}{n\sigma^2} \leq
(1 - 4\sqrt{t'/n}) \right\}\\
&& =pP\left\{ \frac{\|e\|^2_2}{n\sigma^2} \leq (1 - 4\sqrt{t'/n})
\right\} \leq p\exp(-t') \enskip ,
\end{eqnarray*}
and similarly,
\begin{equation*}
P\left\{\max_i \frac{\|e^{(i)}\|^2_2}{n\sigma^2}\geq (1 +
4\sqrt{t'/n}) \right\} \le p\exp(-t')\enskip .
\end{equation*}

We choose $t'$ so that $\mu^2\le C_0 ||e^{(i)}||^2_2/S$ uniformly in
$i$ with probability at least $1-2e^{-pH(r)\epsilon/2}$, so as to
match the rate in Section~A above. Specifically, we set
$t'=\left[pH(r)\varepsilon\right]/2 + \log\left(p/2\right)$.  Hence,
for sufficiently small $\varepsilon$ we have $t' \approx \log(p/2)$.
Therefore, as long as $\mu^2 \leq \left(C_0\sigma^2 n /S\right)
\left[1- 4(t'/n)^{1/2}\right]$, then with probability exceeding $1-
2e^{-pH(r)\varepsilon/2}$, the inequality $\mu^2\le C_0
||e^{(i)}||^2_2/S$ holds uniformly in $i$. Suppose $p = n^{\nu}$,
with $\nu>1$. Under this condition, $t'/n \approx \left(\log_2
n^\nu\right)/n = \nu \left(\log_2 n\right)/n$ and our requirement
thus reduces to $\mu^2 \leq \left(C_0\sigma^2 n/S \right)\left[1-
4\left((\log_2 n)/n\right)^{1/2}\right]$. Similarly, with $t'
\approx \log\left(p/2\right)$, we also have with probability
exceeding $1- 2e^{-pH(r)\varepsilon/2}$ that $\|e^{(i)}\|_2^2 \leq
\sigma^2 n\left[1+4\left((\log_2n)/n\right)^{1/2}\right]$.

Let $\zeta_n^{-}$ and $\zeta_n^{+}$ be defined as in the statement
of Theorem~1.  Then by requiring that $\mu^2\le (C_0\sigma^2
\zeta_n^{-})/S$, from the above results it follows that Condition
(iii) of Lemma 2 holds for all $i=1,\ldots,p$, with high
probability.  Furthermore, with high probability,
$||e^{(i)}||^2_2\le \sigma^2\zeta_n^{+}$, for $i=1,\ldots,p$.
Therefore, by the bound (\ref{eq:zhu.bnd}) in Lemma 2, we have
established the bound (\ref{eq:cs.bound}) in Theorem~1, except for
the constant $C$.

Specifically, it remains for us to establish that $C^{(i)}\le C$ for
all $i$.  Denote the value $C^{(i)}$ for an arbitrary regression by
$\tilde C$.  Note that the $\tilde C$ here in our paper corresponds
to the square of what is called `$C$' in Zhu~\cite{Zhu}. Hence by
equation (17) in \cite{Zhu}, $\tilde C^{1/2}$ is smaller than the
larger root of a quadratic equation of the form
$$
a^2z^2 - (2ac + b)z +(c^2-\tau) = 0 \enskip ,
$$
where $z$ is the argument and $a, b, c$ and $\tau$ are positive
parameters.\footnote{Our notation here is slightly different from
that of~\cite{Zhu}.}  For our purposes, it is enough to remark that
$a > (1 -\delta_{4S} - 2\delta_{5S})/3$, $b$ is bounded by a
constant proportional to $C_0^{1/2}$, $c$ relates to $a$ through the
expression $a=c\left[2\mu S^{1/2}/\zeta^{1/2}\right] - 1-
\delta_{4S}$, and $\tau$ is a constant greater than four,

As the larger root of the above quadratic equation,
$$
\tilde C^{1/2} \leq \frac{2ac+b + \sqrt{(2ac+b)^2 -
4a^2(c^2-\tau)}}{2a^2} \enskip .
$$
Note that $(2ac+b)^2 - 4a^2(c^2-\tau) = 4abc + b^2 +4a^2\tau$, which
is bounded by $ \max( (2a\tau^{1/2} + b)^2 , (2ac + b)^2) $, because
$a, b, c$ and $\tau$ are all positive. Hence we have
\begin{eqnarray}
\tilde C^{1/2} &\leq& \max\left ( \frac{2ac+b + 2a\tau^{1/2} +
b}{2a^2},  \frac{2ac+b
+ 2ac + b}{2a^2}\right) \nonumber \\
&<& \max\left(\frac{\tau^{1/2}+c}{a} + \frac{b}{a^2},\frac{2c}{a} +
\frac{b}{a^
2} \right) \nonumber \\
&<&2\,\frac{\max(c,\tau^{1/2})}{a} + \frac{b}{a^2} \label{eq:a.bnd}
\enskip .
\end{eqnarray}

It remains for us to bound the right-hand side of (\ref{eq:a.bnd}).
Recall that, by construction, $\delta_{4S} + 2\delta_{5S} \leq
\rho(r)$, and that $\rho(r)$ is assumed strictly less than $1$.
Thus, because $b$ is bounded by a term proportional to $C_0^{1/2}$,
the second term in the right-hand side of (\ref{eq:a.bnd}) is
bounded by a term proportional to $(1-\rho(r))^2$.  Furthermore, if
$\tau^{1/2}\ge c$, then the first term is bounded by a term
proportional to $(1-\rho(r))$. Lastly, therefore, suppose that $c >
\tau^{1/2}$ and consider the term
\begin{equation}
\frac{c}{a} = \frac{1}{2\mu S^{1/2}/\zeta^{1/2}}\,
        \frac{a + 1 + \delta_{4S}}{a} \enskip .
\label{eq:c.over.a.bnd}
\end{equation}
The term involving $a$ in the right-hand side of
(\ref{eq:c.over.a.bnd}) is easily bounded, per our reasoning above,
while -- taking, for example, $\mu^2=C_0\zeta/S$ in the condition of
Lemma~\ref{lem:zhu}, the other term is equal to $(4C_0)^{-1/2}$.

Hence, returning to the context of our original problem, for each
$i=1,\ldots,p$, the constant $C^{(i)}$ is bounded by some constant
times $(1-\rho(r))^{-2}$.  Letting $C$ be the largest of these
bounds, our proof of Theorem~1 is complete.


\section{Proof for Theorem 2}
To show that condition (\ref{cond3}) of Theorem~1 holds in the
context of Theorem~2, we first bound the eigenvalue ratio of the
covariance matrix $\Sigma$. For $q\in (0,1)$, the matrix
$\Sigma^{-1}$ is diagonally dominant, and hence, by the
Levy-Desplanques Theorem, non-singular.  Furthermore, since
$\Sigma^{-1}$ is real and symmetric, it is a normal matrix.
Therefore,
$$
\frac{\lambda_{\max}(\Sigma)}{\lambda_{\min}(\Sigma)} =
\frac{1/\lambda_{\min}(\Sigma^{-1})}{1/\lambda_{\max}(\Sigma^{-1})}
= \frac{\lambda_{\max}(\Sigma^{-1})}{\lambda_{\min}(\Sigma^{-1})} =
\kappa_2(\Sigma^{-1}),
$$
where $\kappa_2(\Sigma^{-1})$ is condition number of the precision
matrix $\Sigma^{-1}$. As a result, by an inequality of Guggenheimer,
Edelman, and Johnson~\cite[Pg. 4]{Gug} for condition numbers, we can
bound our eigenvalue ratio as
\begin{equation}
\frac{\lambda_{\max}(\Sigma)}{\lambda_{\min}(\Sigma)} \leq
\frac{2}{|\det(\Sigma^{-1})|}\left(
\frac{\|\Sigma^{-1}\|_F^2}{p}\right)^{p/2} \enskip .
\label{eq:condbound}
\end{equation}

Since $\Sigma^{-1} = I + q D^{-1/2}AD^{-1/2}$, direct calculation
shows that
\begin{eqnarray*}
\|\Sigma^{-1}\|_F^2 &=& p + q^2 \|D^{-1/2}AD^{-1/2}\|_F^2 = p +
q^2\sum_{i\sim j}(\frac{1}{\sqrt{d_id_j}})^2\\
&=& p + q^2 \sum_{i=1}^p\frac{1}{d_i}\sum_{j\sim
i}\frac{1}{d_j}\enskip ,
\end{eqnarray*} where $d_i$ is the \emph{i-th} element of the diagonal
matrix $D$. As for the quantity $|\det(\Sigma^{-1})|$, we note that
\begin{eqnarray*}
|\det(\Sigma^{-1})| &=& |q\det(D^{-1/2})\det(1/qD +A)\det(
D^{-1/2})|\\
&=& q^p \prod_{i=1}^p d_i^{-1}|\det(1/q D + A)| \enskip .
\end{eqnarray*}
Denoting $M = (1/q)D + A$, and applying a result of Ostrowski for
determinants of diagonally dominant matrices (e.g., \cite{ost}), we
find that
\begin{eqnarray*}
|\det(M)| \geq \prod_{i=1}^p(|M_{ii}| +\sum_{j\neq i}|M_{ij}|) &=&
\prod_{i=1}^p(\frac{1}{q}d_i +d_i)\\
&=& \left(\frac{1 + q}{q}\right)^p\prod_{i=1}^p d_i \enskip .
\end{eqnarray*}
Hence, we have $\det(\Sigma^{-1}) \geq (1+q)^p$.

Combining the relevant expressions above, we have that
\begin{eqnarray}
&&\frac{\lambda_{\max}(\Sigma)}{\lambda_{\min}(\Sigma)} \leq
\frac{2}{(1+q)^p}\left( 1+ \frac{q^2 \sum_{i=1}^p \frac{\sum_{j\sim
i}1/d_j}{d_i} }{p}\right)^{p/2} \\
&=&2\left[\frac{1}{(1+q)^2} + (\frac{q}{1+q})^2
\Ave_i\left(HM_{j\sim i}( \frac{1}{d_j})\right) \right]^{p/2}
\enskip . \label{eq:final.cond.bnd}
\end{eqnarray}
Denoting $\eta_1$ and $\eta_2$ as in (\ref{eq:etas}), bounding the
right-hand side of (\ref{eq:final.cond.bnd}) by $\left[1 +
(d_{max}/n)^{1/2}\right] / \left[1 - (d_{max}/n)^{1/2}\right]$, to
enforce the bound (\ref{cond3}), and some trivial manipulation of
the resulting inequality, yields the condition in
(\ref{eq:eta.cond}), as desired.


\section{Proof for Theorem 3}
Note that the difference of the predictor $\hat{\phi}$ and the true
external effect $\phi$ is given as $ \hat\phi - \phi = \tilde{ Y} -
\hat{B}\tilde{Y} - \phi = (I-\hat{B})\tilde{Y} - \phi$.  So the bias
term is just
\begin{eqnarray*}
&& E[(\hat{\phi} - \phi)|\hat{B}] = (I-\hat{B})(I-B)^{-1}\phi - \phi\\
&&=[(I - \hat{B}) - (I - B)](I-B)^{-1}\phi = \Delta (I-B)^{-1}\phi
\enskip ,
\end{eqnarray*}
where $\Delta = B - \hat B$, and hence for the \emph{i-th} component
of the bias term, we have
\begin{eqnarray*}
  \left | E\left[(\hat{\phi}_i -\phi_i)\,\big\vert\,\hat{B}\right]\right |
& = & \left | \Delta_{i\bullet} (I-B)^{-1} \phi \right | \\
& \le & \| \Delta_{i\bullet}\|_2 \quad
        \| (I-B)^{-1} \|_2 \quad
    \| \phi \|_2 \\
& \le & \left[ (C\sigma^2\zeta_n^{+})^{1/2} \,\,
        \lambda^{1/2}_{max}\left((I-B)^{-1}\right)
    \right] \,\, \|\phi\|_2 \enskip .
\end{eqnarray*}
Here the first term in brackets follows from Theorem~1, while the
second follows from the definition of the $\ell_2$ matrix norm.

For the variance of the predictor $\hat{\phi}$, we have
\begin{eqnarray*}
&&\Var\left[\hat{\phi}\mid\hat{B}\right] = (I-\hat{ B})(I-B)^{-1}(I - \hat{ B})^T \sigma^2\\
&=& (I-B + B - \hat{ B})(I-B)^{-1}(I-B + B - \hat{ B})^T \sigma^2\\
& = & (I-B)\sigma^2 + \left[ \Delta (I-B)^{-1} \Delta^T +
2\Delta\right]
    \sigma^2\enskip .
\end{eqnarray*}
So the variance of the $i$th element $\hat\phi_i$ of $\hat{\phi}$ is
\begin{equation*}
\Var\left[\hat{\phi}_i \,\big\vert\, \hat{B}\right] = \sigma^2 +
    \Delta_{i\bullet} (I-B)^{-1} \Delta_{i\bullet}^T \sigma^2 \enskip ,
\end{equation*}
since $B_{ii}=\hat B_{ii} = 0$.  The absolute value of the second
term is bounded by $\left(C\sigma^2 \zeta_n^{+}\right)
            \lambda_{max}\left((I-B)^{-1}\right)$,
and thus (\ref{eq:phihat.var.bnd}) follows.

\section{Proof for Lemma 1.}

Under the model $\Sigma^{-1} = I+ qD^{-1/2}AD^{-1/2}$, the largest
number of non-zero entries $S$ is $d_{\max}$, the maximal degree of
the network. So by the eigenvalue condition in Theorem~1, we have
\begin{equation*}
\lambda_{\max}((I-B)^{-1}) \leq \lambda_{\min}((I-B)^{-1})\cdot
\left(\frac{1 +
\sqrt{d_{\max}/n}}{1-\sqrt{d_{\max}/n}}\right)^2\enskip .
\end{equation*}
%
Now $\lambda_{\min}((I-B)^{-1}) = \lambda^{-1}_{\max}((I-B)) =
\lambda^{-1}_{\max}(I+qD^{-1/2}AD^{-1/2})$. And clearly
$\lambda_{\max}(I+qD^{-1/2}AD^{-1/2}) = 1 +
\lambda_{\max}(qD^{-1/2}AD^{-1/2})$.  Furthermore,
\begin{eqnarray*}
&&\lambda_{\max}(D^{-1/2}AD^{-1/2}) =
    \max_{ x }\, \frac{x'D^{-1/2}AD^{-1/2}x}{x'x}\\
&=&\max_{ x }\, \left[\,
    \frac{(D^{-1/2}x)'A(D^{-1/2}x)}{(D^{-1/2}x)'(D^{-1/2}x)}\cdot
\frac{(D^{-1/2}x)'(D^{-1/2}x)}{x'x}\,\right] \\
&\geq&  \lambda_{\max}(A) \lambda_{\min}(D^{-1})
=\frac{\lambda_{\max}(A)}{d_{max}}
\end{eqnarray*}
By~\cite[Lem. 8.6]{Chung}, $\lambda_{\max}(A) > d^{1/2}_{\max}$.
Therefore, $\lambda_{\max}(I+qD^{-1/2}AD^{-1/2}) \geq 1+q
d^{-1/2}_{\max}$.

Combining these results, we have
\begin{eqnarray*}
\lambda_{\max}((I-B)^{-1}) &\leq&
\frac{1}{1+q\frac{1}{\sqrt{d_{\max}}}}\cdot \left(\frac{1 +
\sqrt{d_{\max}/n}}{1-\sqrt{d_{\max}/n}}\right)^2\\
&=& \frac{\sqrt{d_{\max}}}{q + \sqrt{d_{\max}}}\cdot \left(\frac{1 +
\sqrt{d_{\max}/n}}{1-\sqrt{d_{\max}/n}}\right)^2 \enskip .
\end{eqnarray*}



\bigskip\bigskip

\section*{Acknowledgment}

\ifCLASSOPTIONcaptionsoff
  \newpage
\fi

\end{document}